\documentclass[useAMS,usenatbib]{mn2e}

\usepackage{psfig}
\usepackage[dvips]{graphicx}
\usepackage{lscape}

\def\lesssim{\mathrel{\hbox{\rlap{\hbox{\lower4pt\hbox{$\sim$}}}\hbox{$<$}}}}
\def\gtrsim{\mathrel{\hbox{\rlap{\hbox{\lower4pt\hbox{$\sim$}}}\hbox{$>$}}}}
\newcommand{\kms}{km\,s$^{-1}$}
\newcommand{\Msol}{M$_\odot$}
\newcommand{\Teff}{$T_{\mbox{\scriptsize eff}}$}
\newcommand{\Feh}{\mbox{$\mbox{[Fe/H]}$}}
\newcommand{\Zh}{\mbox{$\mbox{[Z/H]}$}}

\title[Extended MILES stellar population models]{MIUSCAT: extended MILES spectral coverage. I. Stellar populations
synthesis models.}
\author[Vazdekis et al.]{A. Vazdekis$^{1,2}$\thanks{E-mail: vazdekis@iac.es}, 
    E. Ricciardelli$^3$, A.J. Cenarro$^4$, 
J. G. Rivero-Gonz\'alez$^5$, 
\newauthor L. A. D\'{\i}az-Garc\'{\i}a$^4$, J. Falc\'on-Barroso$^{1,2}$ \\
$^{1}$Instituto de Astrof{\'{\i}}sica de Canarias, V{\'{\i}}a Lactea s/n, E-38200 La Laguna, Tenerife, Spain\\
$^{2}$Departamento de Astrof{\'{\i}}sica, Universidad de La Laguna, E-38205, Tenerife, Spain\\
$^{3}$Departament d'Astronomia i Astrofisica, Universitat de Valencia, c/ Dr. Moliner 50, E-46100 - Burjassot, Valencia, Spain\\
$^{4}$Centro de Estudios de F{\'{\i}}sica del Cosmos de Arag\'on, Plaza de San Juan 1, Planta 2, E-44001, Teruel, Spain\\
$^{5}$Universit\"atssternwarte M\"unchen, Scheinerstr. 1, 81679 M\"unchen, Germany 
}

\begin{document}
\date{Accepted ...  Received ...; in original form ...
}
\pagerange{\pageref{firstpage}--\pageref{lastpage}} \pubyear{...}
\maketitle
\label{firstpage}

\begin{abstract}
We extend the spectral range of our stellar population synthesis models based
on the MILES and CaT empirical stellar spectral libraries. For this purpose we
combine these two libraries with the Indo-U.S. to construct composite stellar
spectra to feed our models. The spectral energy distributions (SEDs) computed
with these models and the originally published models are combined to construct
composite SEDs for single-age, single-metallicity stellar populations (SSPs)
covering the range $\lambda\lambda$ 3465 -- 9469\,\AA\ at moderately high, and
uniform, resolution (FWHM$=$2.51\AA). The colours derived from these SSP SEDs
provide good fits to Galactic globular cluster data. We find that the colours
involving redder filters are very sensitive to the IMF, as well as a number of
features and molecular bands throughout the spectra. To illustrate the
potential use of these models we focus on the Na{\sc I} doublet at 8200\,\AA\
and with the aid of the newly synthesized SSP model SEDs we define a new
IMF-sensitive index that is based on this feature, which overcomes various
limitations from previous index definitions for low velocity
dispersion stellar systems.  We propose an index-index diagram based on 
this feature and the neighboring Ca{\sc II} triplet at 8600\,\AA,
to constrain the IMF if the age and {\mbox{$\mbox{[Na/Fe]}$}} abundance are 
known.  Finally
we also show a survey-oriented spectrophotometric application which evidences
the accurate flux calibration of these models for carrying out reliable
spectral fitting techniques. These models are available through our
user-friendly website.
\end{abstract}


%%%%%%%%%%%%%%%%%%%%%%%%%%%%%%%%%%%%%%%%%%%%%%%%%%%%%%%%%%%%%%%%%%%%%%%%%%%%%%%%

\section{Introduction}
\label{intro}

A quantitative study of the stellar content of galaxies and star clusters
requires the use of the so called stellar population synthesis models. The
method consists in comparing observables like colours, line-strength indices or
full spectral energy distributions (SEDs) to the predictions of these models
(e.g., \citealt{Tinsley80}). Furthermore these models are widely employed in many other
type of studies such as, e.g., to provide the necessary templates for kinematic
or redshift measurement or to determine galaxy stellar masses with the aid of
the mass-to-light ratios (M/L). Most recent models employ at least three main
ingredients, which determine the quality of the predictions: a prescription for
the initial mass function (IMF), a set of stellar evolutionary isochrones and
stellar spectral libraries. Galaxy observables are predicted by adding the
contributions of all possible stars, in proportions prescribed by stellar
evolution models. The required SEDs, colours or absorption line-strengths of all
these stars are obtained from the stellar libraries. 

Colours, absorption line-strengths and other observables from these models have
been widely used in the literature for interpreting the integrated light of
stellar clusters and galaxies. In the recent years, it has become
common practice to use
model SEDs, mainly for single-burst stellar populations (SSPs), at moderately
high spectral resolution. These models provide new means for lifting the main
degeneracies hampering stellar populations studies, such as the one between the
age and the metallicity (e.g., \citealt{Worthey94}). For example these SEDs have
allowed us to define new indices with enhanced abilities to disentangle the age
from the metallicity (e.g., \citealt{VazdekisArimoto99, Cervantes09}), or even the
IMF (e.g., \citealt{Schiavon00, Vazdekis03}). In recent years it is becoming
increasingly popular to use these model SEDs to derive relevant stellar
population parameters, including the Star Formation History (SFH), by means of
full spectrum-fitting approach (e.g., \citealt{Cid05, Koleva09, Tojeiro11}), which
represents an alternative to the more standard approach of fitting a selected
number of line-strength indices (see, e.g., \citealt{Trager98}, and references
therein). Furthermore these model SEDs have been used in a variety of
applications. For example, these SSP SEDs have been shown to improve the
analysis of galaxy kinematics for both absorption and emission lines (e.g.,
\citealt{FalconBarroso03, Sarzi06}).  

There are models that are fed with theoretical stellar atmospheres (e.g.,
\citealt{Schiavon00, GonzalezDelgado05, Coelho07}), whereas other models employ
empirical libraries, mostly in the visible (e.g., \citealt{Vazdekis99, BC03,
LeBorgne04, Conroy10, Vazdekis10, Maraston11, Conroy11}), but also in other
spectral ranges (e.g., \citealt{Vazdekis03, Conroy11, Maraston11}). Models based
on empirical libraries are free from the uncertainties in the underlying model
atmospheric calculations and tend to provide good fits to both, absorption
line-strengths and spectra (e.g. \citealt{Vazdekis10, Maraston11} and photometric
data  (e.g., \citealt{Maraston09, Peacock11}). Unlike the models based on
theoretical libraries, these models are generally restricted in their SSP
parameter coverage, as most of the stars come from the solar neighbourhood, with
e.g. a limited metallicity and elemental abundance ratio coverage. The quality
of the resulting model SEDs relies to a great extent on the atmospheric
parameters (temperature, gravity and metallicity) coverage of the library. 

We published model SEDs in the near-IR at 1.5\,\AA\ (FWHM) resolution
\citep{Vazdekis03} and in the optical range \citep{Vazdekis10} at
FWHM$=$2.51\,\AA\ \citep{FalconBarroso11}, which employ the CaT and MILES
empirical stellar spectral libraries of \citet{Cenarro01} and \citet{Sanchez06},
respectively. These libraries show improved coverage of the stellar parameters
and good flux-calibration quality among other characteristics. Unfortunately
there is a spectral gap around 8000\,\AA, of about 900\,\AA, which is not
covered between these two model SED predictions. This gap prevents us to derive
a variety of widely employed colours, from $U$ to $I$ broad band filters, and to
exploit potentially interesting  absorption features in that spectral region,
such as the Na{\sc I} at 8200\,\AA, which has been shown to be useful for
constraining the dwarf/giant ratios in galaxies (e.g. \citealt{Faber80}).

This is the first paper of a series where we present the extended spectral
coverage of the SEDs predicted by our stellar population synthesis models. For
this purpose we make use of the Indo-U.S. stellar spectral library
\citep{Valdes04} to fill-in this gap and to extend blueward and redward the
spectral coverage of our model SEDs. These SEDs constitute an extension of our
"base models", as we combine scaled-solar isochrones with empirical stellar
spectral libraries, which follow the chemical evolution pattern of the solar
neighbourhood. The models rely as much as possible on empirical ingredients, not
just on the stellar spectra, but also on extensive photometric libraries, which
are used to determine the transformations from the theoretical parameters of the
isochrones to observational quantities. Our predicted SEDs are therefore
self-consistent, and scaled-solar for solar metallicity. In the low
metallicity regime, however, our models combine scaled-solar isochrones with
stellar spectra that do not show this abundance ratio pattern (e.g.,
\citealt{Edvardsson93, Schiavon07}). An extensive comparison of the predicted
photometric properties with data of globular clusters of the MW and M\,31 and
early-type galaxies from the SDSS is shown in a second paper (Ricciardelli et
al. 2012; hereafter Paper~II). 

The paper is organized as follows. In Section 2 we show the main ingredients of
our models, including the composite stellar spectral library that we constructed
to feed these models. In Section 3 we describe the reliability and quality of
the resulting models and the behaviour of the  MIUSCAT SSP SEDs and colours
derived from these spectra. Section 4 illustrates with various examples the
potential use of these models. In Section 5 we describe the web-based
user-friendly facilities to download and handle these model SEDs. Finally in
Section 6 we summarize our results. 

%%%%%%%%%%%%%%%%%%%%%%%%%%%%%%%%%%%%%%%%%%%%%%%%%%%%%%%%%%%%%%%%%%%%%%%%%%%%%%%%

\section{Models}
\label{sec:models}

The MIUSCAT SSP SEDS presented here represent an extension of the models
published in \citet{Vazdekis03} and \citet{Vazdekis10} to cover the whole
spectral range $\lambda\lambda$ 3465 - 9469\,\AA. We provide here a brief
summary of the main ingredients employed by these models and focus on the new
addition that allows us to achieve such spectral range extension,  i.e. the
MIUSCAT stellar spectral library.

\subsection{Main ingredients}
\label{sec:ingredients}

We employ the solar-scaled theoretical isochrones of \citet{Girardi00}, which
cover a wide range of ages, well sparsed from 0.063 to 17.8\,Gyr, and six
metallicity bins ($Z=$0.0004, 0.001, 0.004, 0.008, 0.019 and 0.03), where 0.019
represents the solar value. The isochrones include the latest stages of the
stellar evolution, including a simple synthetic prescription for incorporating
the thermally pulsing AGB regime to the point of complete envelope ejection. The
range of initial stellar masses extends from 0.15 to 7${\rm M}_{\odot}$. The
input physics of these models was updated with respect to \citet{Bertelli94}
with an improved version of the equation of state, the opacities of
\citet{Alexander94} and a milder convective overshoot scheme. A helium fraction
was adopted according to the relation: $Y\approx0.23+2.25Z$. 

We use the theoretical parameters of the isochrones (\Teff, $\log g$, \Zh) to
obtain stellar fluxes on the basis of empirical relations between colours and
stellar parameters (temperature, metallicity and gravity), instead of using
theoretical stellar atmospheres. We mostly use the metallicity-dependent
empirical relations of \citet{Alonso96, Alonso99}; respectively, for dwarfs and
giants. Each of these libraries are composed of $\sim$~500 stars and the
obtained temperature scales are based on the IR-Flux method, i.e. only
marginally dependent on model atmospheres. We use the empirical compilation of
\citet{Lejeune97, Lejeune98} (and references therein) for the coolest dwarfs (
$T_{\mbox{\scriptsize eff}}\le4000\,K$) and giants ($T_{\mbox{\scriptsize
eff}}\le3500\,K$) for solar metallicity, and also for stars with temperatures
above $\sim$8000\,K. For these low temperatures we use a semi-empirical approach
to other metallicities on the basis of these relations and the  model
atmospheres of \citet{Bessell89, Bessell91} and the library of \citet{Fluks94}.
We also employ the metal-dependent bolometric corrections given by
\citet{Alonso95, Alonso99} for dwarfs and giants, respectively. We adopt
$BC_{\odot}=-0.12$ and a bolometric magnitude of 4.70 for the Sun.

Several IMFs are considered: the two power-law IMFs described in
\citet{Vazdekis96} (i.e unimodal and bimodal), the two characterized by its
slope $\mu$ as a free parameter, and the multi-part power-law IMFs of
\citet{Kroupa01} (i.e. universal and revised). The \citet{Salpeter55} IMF is
obtained by adopting the unimodal case with slope $\mu=1.3$. An extensive
description of these definitions is given in V03 (Appendix~A).  We set the lower
and upper mass-cutoff of the IMF to 0.1 and 100\,\Msol, respectively.  

\subsection{MIUSCAT stellar spectral library}
\label{sec:MIUSCATlib}

To extend the spectral coverage of our MILES-based models \citep{Vazdekis10} we
employ three empirical stellar libraries at moderately-high spectral resolution:
MILES \citep{Sanchez06} and the near-IR CaT library of \citet{Cenarro01}, both
with very good relative flux-calibration, and the Indo-U.S.
\footnote{http://www.noao.edu/cflib/} \citep{Valdes04}, which covers a
significantly wider spectral range. This library allows us to fill-in the gap
left between the MILES and CaT libraries and to extend the wavelength coverage
slightly blueward of MILES and redward of the CaT library. 

MILES includes 985 stars covering the range 3525-7500\,\AA\ with a spectral
resolution of 2.5\,\AA (FWHM), as recently shown in \citet{FalconBarroso11}.
The CaT library \citep{Cenarro01} consists of 706 stars in the range
8350-9020\,\AA\ at resolution 1.5\,\AA. The INDO-U.S. library includes $\sim$1200
stars with spectra that typically cover from 3465 to 9469\,\AA. In
\citet{FalconBarroso11} we also have recently shown that the spectral resolution
of this library is slightly larger (FWHM$=$1.36\,\AA) than the approximate
value provided in the original Indo-U.S. paper (FWHM$\sim$1.2\,\AA).

The atmospheric parameters of all the Indo-U.S. stars in common with MILES
and/or CaT libraries have been compared in order to calculate the necessary
transformations for obtaining an homogenized set of stellar parameters with
MILES. From this comparison we obtained 

\begin{equation}
{\rm Teff}_M = 1.02{\rm Teff}_{IU} - 108.31
\end{equation}

\begin{equation}
\log {\rm g}_M = 1.0467\log {\rm g}_{IU} - 0.1077
\end{equation}

\begin{equation}
{\rm[Fe/H]}_M ={\rm[Fe/H]}_{IU}-0.02
\end{equation}

\noindent As it can be seen, a linear correction was required to bring the
published Indo-U.S. temperatures and gravities onto the MILES/CaT system,
whereas the metallicities were matched by applying a small offset.

We selected the Indo-U.S. stars whose  spectra cover the desired wavelength
range without gaps.  We ended up with 574 stars, from which 142 have been
removed due to the presence of strong telluric absorption residuals in their
spectra. The final catalog is composed of 432 stars. Finally we flagged 76 of
these stars for showing moderate telluric residuals.

In order to obtain a constant resolution and dispersion value within the desired
spectral range (3465 to 9469\,\AA) for the composite MIUSCAT spectra, we first
convolved the CaT and Indo-U.S. spectra with the required gaussian kernels to
match the lower resolution of MILES (FWHM$=$2.5\,\AA). The next step consisted
in resampling the smoothed spectra to match the dispersion of MILES (0.9\,\AA).

Unlike the MILES and CaT libraries the flux-calibration of the Indo-U.S. spectra is
not very good as a narrow slit was employed during the observations to achieve a
moderately high spectral resolution (see \citealt{Valdes04} for
details). 
%This
%extent is further evident from the comparison shown in Fig.~\ref{VIstars}. Here
%the Johnson-Cousin filter responses were convolved with the spectra of the
%Indo-U.S. stars to obtain the $V-I$ colours, which are then compared to the
%expected values according to the empirical photometric relations described in
%Section~\ref{sec:ingredients} that allow us to transform the theoretical
%parameters of the isochrones to the observational plane.

We use the Indo-U.S. spectra to fill-in the gap left between MILES and CaT
spectral ranges, and also to extend toward the blue and red the wavelength
coverage of MILES and CaT libraries, respectively. For each of the selected
Indo-U.S. stars we use the MILES and/or CaT spectra, when available, and then
plug the relevant Indo-U.S. spectral ranges to construct the composite MIUSCAT
stellar spectrum. We find 208 stars in common to these three libraries. There
are 150 Indo-U.S. stars in common with MILES and 114 in common with
CaT. If a star is not present in either MILES or CaT, or even in these two
libraries, we use the interpolators described in \citet{Vazdekis10} and
\citet{Vazdekis03} to synthesize the corresponding MILES and CaT spectra,
respectively.

To join the stellar spectra we first calibrate the Indo-U.S. spectrum in
the region of the gap present between MILES and CaT.  The relative flux scale
of the MILES and CaT spectra is set by forcing the spectra to have
the $V-I$ color corresponding to its stellar parameters, following the
method described in V03 (Section 3.2.1). The $V-I$ color is derived from
the photometric relations described in Section~\ref{sec:ingredients}.  Note
that this method ensures a consistent approach throughout the various steps of
the model computations.  The Indo-U.S. spectrum in the gap is forced to overlap
the MILES and CaT spectra at their red and blue edges, respectively. This is
performed by approximating the continuum shape of the  original and re-scaled
Indo-U.S. spectra in the gap with a straight line joining the points at
$\lambda\lambda$ 7360-7385\,\AA and $\lambda\lambda$ 8390-8415\,\AA (see
Table~\ref{tab:matching_bands}).  The Indo-U.S. spectrum is then scaled
according to the ratio between the flux values on these two straight lines on
each of the pixels.  In the  Indo-U.S. edges, i.e. bluewards MILES and redwards
CaT, the INDO-U.S. spectrum is simply re-scaled to match the MILES and CaT
continua.  We scale the relevant Indo-U.S.  spectral range by matching the flux
of the MILES and CaT spectra in well selected narrow bands located at nearly
the edges of spectral ranges of these libraries. These overlapping bands are
chosen to be sufficiently wide to achieve the statistics. However these bands
are sufficiently narrow to avoid prominent absorption features and to enable to
use most of the MILES and CaT spectral ranges. The limiting wavelengths of the
selected overlapping wavelength regions are listed in
Table~\ref{tab:matching_bands}. 

Finally as the limiting wavelengths of the spectral range covered by each
stellar spectrum of the Indo-U.S. library differs slightly from each other we
consider a common range for all the stars ($\lambda\lambda$ 3464.9 --
9468.8\,\AA).

\begin{table}
\centering
\caption{Adopted merging wavelength ranges for matching the stellar spectra of the Indo-U.S., MILES
and CaT libraries.\label{tab:matching_bands}} 
\begin{tabular}{lc}
\hline
 Libraries        & Overlap wavelength range (\AA) \\
\hline
Indo-U.S.,MILES  &$ 3540-3564$\\
MILES,Indo-U.S &$ 7360-7385$\\
Indo-U.S,CaT  & $8390-8415$\\
CaT,Indo-U.S & $8926-8950$\\
\hline
\end{tabular}
\end{table}

\begin{figure}
\includegraphics[angle=0,width=\columnwidth]{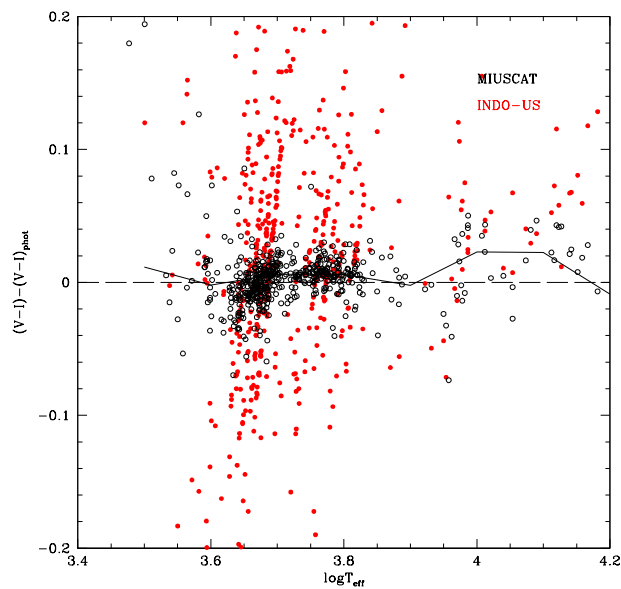}
\caption{We plot against $\log$\Teff\ the residuals obtained by comparing the
synthetic ($V-I$) colour to the expected colour value ($(V-I)_{\rm phot}$) for the
MIUSCAT stars. The latter values are obtained by applying the empirical
photometric relations described in Section~\ref{sec:ingredients} according to
the stellar parameters. The synthetic colours are obtained by convolving the
resulting MIUSCAT (black open symbols) and the original Indo-U.S.(red solid
symbols) stellar spectra with the filter responses (Johnson-Cousin system). Note
the large dispersion obtained with the original Indo-U.S. spectra due to the
limited flux-calibration quality (see the text for details).  The solid
black line represents the median of the MIUSCAT residuals for equally spaced
logarithmic temperature interval.} 
\label{VIstars}
\end{figure}

\begin{figure}
\includegraphics[angle=0,width=\columnwidth]{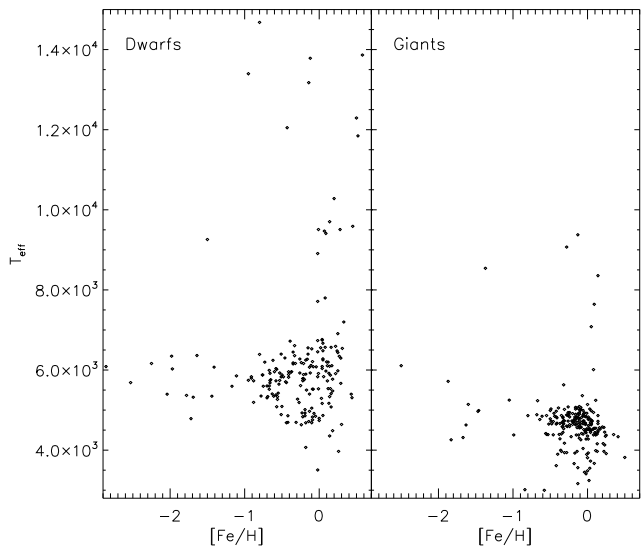}
\caption{The fundamental parameter coverage of the MIUSCAT stars is shown for
dwarfs (left panel) and giants (right panel).}
\label{miusc}
\end{figure}

In Fig.~\ref{VIstars} we show the residuals for the synthetic $V-I$ colours of
both, the resulting MIUSCAT and the original Indo-U.S. spectra, compared to the
expected colour values as a function of temperature. The latter are obtained by
applying the empirical photometric relations described in
Section~\ref{sec:ingredients}, which are based on extensive photometric
libraries, according to the stellar parameters. The synthetic colours are
obtained by convolving the stellar spectra with the filter responses
(Johnson-Cousin system). As expected the original Indo-U.S. spectra show large
residuals due to the limited flux-calibration quality. The composite MIUSCAT
stellar spectra show good agreement with the expected photometric colours.
Indeed for the vast majority of the stars we obtain residuals smaller than the
photometric zero-point (i.e. $\sim$0.02\,mag).

\subsubsection{Stellar atmospheric parameter coverage}
\label{sec:coverage}

Figure\,\ref{miusc} shows the parameter coverage of the MIUSCAT stars for dwarfs
and giants (separated at $\log g = 3$). The parameters of the MIUSCAT stars
plotted here are on the MILES system, which is the result of an extensive
compilation from the literature, homogenized by taking as a reference the stars
in common with \citet{Soubiran98} (see \citealt{Cenarro07}).
Fig.~\ref{miusc} shows that, at solar metallicity, all types of stars are well
represented. For metallicities lower than $\Feh \sim -0.8$, however, the
coverage is rather poor and therefore no reliable models can be computed. For
the metallicity range between solar and $\Feh \sim -0.8$ the coverage of both
dwarfs and giants is good, with the notably exception of dwarfs with
temperatures above $\sim$7000\,K. As these stars may represent the hotter Main
Sequence, including the Turn Off, the resulting model predictions will be of
lower quality in comparison to those for the older ages. Finally the coverage of
metal-rich dwarf and giant stars allows us to safely compute SSP SEDs for $\Zh
\sim 0.2$, including for the young and intermediate age regimes. A quantitative
analysis of the quality of the models based on the atmospheric parameter
coverage of MIUSCAT is provided in Section~\ref{sec:synthesis}.  

\subsubsection{U magnitude}
\label{sec:u}

Since the bluest wavelength of the MIUSCAT spectra is 3464.9\,\AA, we cannot
accurately measure the flux in the Johnson $U$ band, as this filter starts at
$\approx$3050\,\AA\  \citep{Buser78}. A similar situation applies to the SDSS
$u$ filter. As these filters are commonly used in the literature and the MIUSCAT
spectral range provides most, but not all, the flux in these bands we have
estimated the required corrections to be able to predict these quantities from
the resulting MIUSCAT SSP model SEDs.  To estimate the missing flux bluewards
3464.9\,\AA\ we employ the \citet{Pickles98} library of stars, which covers the
spectral range $\lambda\lambda$ 1150-25000\,\AA\ at low resolution. Note that
this library is mainly composed by solar metallicity stars. We
estimate the missing fluxes by convolving the Pickles spectra with the filter
responses of these bands to obtain both the total flux and the flux in the
wavelength range covered by MIUSCAT, i.e, redwards 3464.9\,\AA. 

\begin{figure}
\includegraphics[angle=0,width=\columnwidth]{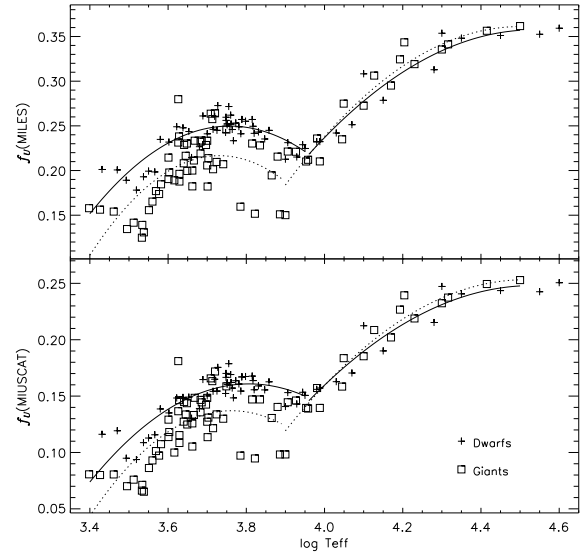}
\caption{Missing flux correcting factor for the Johnson U filter versus
$\log$\Teff\ for dwarfs (crosses) and giants (open squares). For this purpose we
convolved the U filter response with the low-resolution stellar spectra of the
\citet{Pickles98} library. In the upper panel we show the missing flux factor
obtained for the MILES spectral range, whereas in the lower panel we show the
one corresponding to the MIUSCAT range. Note that the required correction for
the MIUSCAT range is significantly smaller. The fits for dwarfs (solid lines)
and giants (dotted lines) are given in Eq.~\ref{eq:fU}.}
\label{fU}
\end{figure}

The correction factors, given by the missing flux divided by the total
flux, for the Johnson $U$ filter, $f_U$,
are shown in Fig.~\ref{fU} for the different spectral types by adopting the two,
the limiting wavelengths of MILES (upper panel) and MIUSCAT (lower panel).
Despite the fact that MIUSCAT only extends to the blue 75\,\AA\ the MILES
spectral range, the missing flux is reduced by about $\sim$10\%. This is not
surprising given the rapid decay of the $U$ filter transmission at these
wavelengths. As expected hot stars require larger corrections, particularly
those with temperatures above $\sim$10000\,K, whilst stars with temperatures
below $\sim$4000\,K show corrections factors smaller than 10\%. The flux
corrections are in the range 10 -- 15\% for the stars that mainly contribute to
the old stellar populations. 
We obtain separate fits for dwarfs and giants in the MIUSCAT spectral
range as follows

\begin{equation}
 \begin{array}{r@{}c@{}l@{}@{}l@{}l@{}}
f_{U} & = & -7.4540 + 4.0006 T -0.5254 T^2 & ,\, T<3.955,  & \,dwarfs\\
f_{U} & = & -6.1348 + 2.8159 T -0.3106 T^2 & ,\, T>3.955,  & \,dwarfs\\
\\
f_{U} & = & -9.9324 + 5.3528 T -0.7114 T^2 & ,\, T<3.895,  & \,giants\\
f_{U} & = & -7.0157 + 3.2215 T -0.3569 T^2 & ,\, T>3.895,  & \,giants\\
 \end{array}
 \label{eq:fU}
\end{equation}

\noindent where $T$ means $\log$\Teff. The total $U$  magnitude is derived by
measuring the flux of the MIUSCAT spectrum through the U filter redwards
3464.9\,\AA and correcting it by the missing flux. These fractions are taken
into account during the integration along the isochrone to obtain the
corresponding missing flux corrections for the SSP. It is worth noting that these
corrections are strictly valid for solar metallicity due to the limited coverage
of the \citet{Pickles98} library in this parameter. However the main
contribution to this correction factor comes from the missing spectral range,
which represents a small fraction of the total flux in the $U$ band.

\subsection{MIUSCAT SSP SEDs}
\label{sec:synthesis}

We implement the MIUSCAT stellar library in the models as described in
\citet{Vazdekis10}. The spectra of the stars are integrated  along the isochrone
taking into account their number per mass bin according to the adopted IMF. For
this purpose, each requested stellar spectrum is normalized to the corresponding
flux in the V band, following our prescription for converting the theoretical
parameters of the isochrones to the observational plane, as described in
Section~\ref{sec:ingredients}. Note that the latter is performed on the basis of
empirical photometric stellar libraries, rather than relying on theoretical
stellar atmospheres, as it is usually done in other stellar population synthesis
codes. Therefore the SSP SED, $S_{\lambda}(t,{\rm [Z/H]})$, is computed as
follows:

\begin{eqnarray}
S_{\lambda}(t,{\rm [Z/H]})&=&\int_{m_{\rm l}}^{m_{\rm t}}
S_{\lambda}(m,t,{\rm [Z/H]})N(m,t)\times \nonumber \\
&& {F_V}(m,t,{\rm [Z/H]})dm,
\label{eq:SSP}
\end{eqnarray}

\noindent  where $S_{\lambda}(m,t,{\rm [Z/H]})$ is the empirical spectrum,
normalized in the V band, corresponding to a star of mass $m$ and metallicity
[Z/H], which is alive at the age assumed for the stellar population $t$.
$N(m,t)$ is the number of this type of star, which depends on the adopted IMF.
$m_{\rm l}$ and $m_{\rm t}$ are the stars with the smallest and largest stellar
masses, respectively, which are alive in the SSP. The upper mass limit depends
on the age of the stellar population. Finally, ${F_V}(m,t,{\rm [Z/H]})$ is its
flux in the V band. The absolute flux scaling scheeme is described in
\citet{FalconBarroso11}.

The spectrum of each requested star, with a given set of atmospheric parameters,
is synthesized according to the interpolating algorithm described in
\citet{Vazdekis03} and \citet{Vazdekis10}. This code finds the closest stars and
weight them according to the distance to the requested point ($\theta_{0}$,
$\log g_{0}$, [Fe/H]$_{0}$) in the stellar parameter space and the
signal-to-noise of their spectra. The method is optimized to minimize the errors
in case of gaps and asymmetries in the distribution of stars around the
requested point. For a full description of the algorithm we refer the reader to
\citet{Vazdekis03} (Appendix B).

We also take into account during the integration along the isochrone the missing
fraction of the total flux in the $U$ broad band, i.e. bluewards 3464.9\,\AA,
for each star, $f_{U_i}$, which is calculated according to Eq.~\ref{eq:fU}. The
resulting fraction for the SSP, $f_{U_{SSP}}$, is then used to correct the flux
obtained by convolving the MIUSCAT SSP SED with the response of the $U$ filter.
As a test of consistency of these filter measurements, we computed the
$f_{U_{SSP}}$ fractions obtained from the SSP SEDs of \citet{BC03} and
\citet{Maraston05} and applied these corrections to the fluxes measured in the
MIUSCAT SEDs. The magnitudes computed in the two ways are fully consistent, with
difference of the order of 0.01\,mag for all the metallicities. This supports
our approach of applying the correction factors derived from the solar
metallicity stars of the \citet{Pickles98} library as discussed in
Section~\ref{sec:u}.

\begin{figure}
\includegraphics[angle=0,width=\columnwidth]{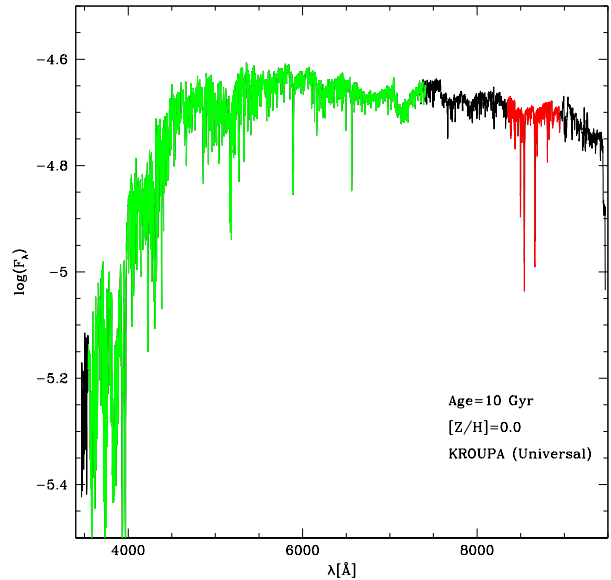}
\caption{An illustrative example of a MIUSCAT SED corresponding to an SSP of
solar metallicity and 10\,Gyr, computed with Kroupa Universal IMF. The joined
spectra are shown in different colours: green for MILES, red for CaT and black
for the Indo-U.S. Note that in the MILES and CaT spectral regions the SEDs are
identical to the ones published on the basis of these two libraries, except the
fact that the CaT model SED has been smoothed to 2.51\,\AA\ (FWHM).}
\label{SSPspectrum}
\end{figure}

Once an SSP SED based on the MIUSCAT library has been computed we combine it
with the original MILES and CaT SSP SEDs of the same age, metallicity and IMF,
to construct the final MIUSCAT model SED. For this purpose we follow the same
approach described for constructing the composite MIUSCAT stars (see
Section~\ref{sec:MIUSCATlib} to recalibrate the models.  This approach
allows us to keep the MILES and CaT SSP SEDs, which are computed on the basis
of these two libraries, untouched (except the smoothing applied to the CaT SED
to match the lower resolution of MILES, i.e. 2.51\,\AA). Although the SSP SEDs
computed on the basis of the composite MIUSCAT stars can be safely used in the
MILES and CaT spectral ranges, our approach warranties a significantly higher
quality in these two ranges as we rely on model computations involving a larger
number of stars with better atmospheric coverage. Therefore these new SEDs,
which are computed on the basis of the composite MIUSCAT stars, are only used
for the wavelength regions that are not covered by the MILES and CaT spectral
ranges. In summary, the MILES and CaT spectral ranges contained in the final
MIUSCAT SSP SEDs are the ones of \citet{Vazdekis03} and \citet{Vazdekis10}
SEDs, as updated in this work.  An example of a composite MIUSCAT SSP SED is shown in
Fig.~\ref{SSPspectrum}. Table~\ref{tab:SEDproperties} summarizes the spectral
properties of these models and the SSP parameters for which our predictions can
be safely used.  The models assume a total initial mass of 1\,M$_{\odot}$.

\begin{figure}
\includegraphics[angle=0,width=\columnwidth]{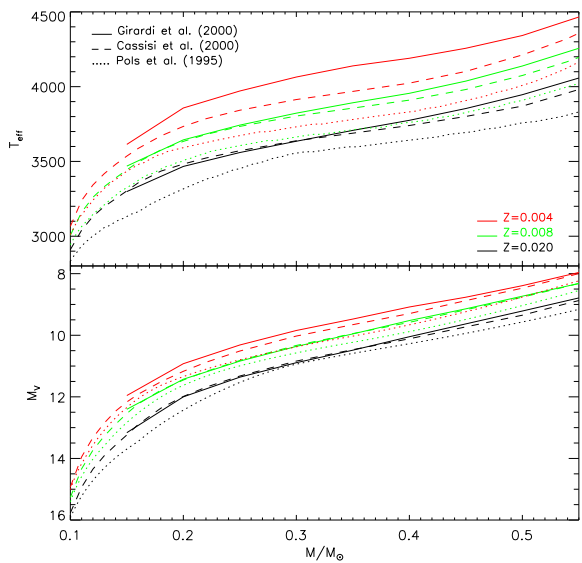}
\caption{Effective temperatures (upper panel) and V band magnitudes
(lower panel) of very low Main Sequence stars for three different set of models
and three metallicities, as indicated in the insets of the upper panel.}
\label{VLMS}
\end{figure}

For unimodal IMFs with very steep slopes ($\mu > 1.3$) the very low MS stars
with $M<0.5$\,\Msol\ might have a noticeable impact on the resulting model
SEDs, particularly in the redder wavelength ranges (e.g. \citealt{Vazdekis96,
Conroy11}).
In Fig.~\ref{VLMS} we show the temperatures and V band magnitudes for three
different sets of models: the \citet{Girardi00} that are employed in this work,
the ones of \citet{Cassisi00} that will be implemented in a forthcoming paper,
and those of \citet{Pols95} that were employed in \citet{Vazdekis96} models.
Note that the models of \citet{Girardi00} are the hottest ones. Furthermore the
obtained discrepancy in temperatures increases with decreasing metallicity. The
difference in temperatures between the hottest, i.e. \citet{Girardi00} and the
coolest models, i.e. \citet{Pols95}, are rather similar to those reported in
\citet{An09}, when comparing the low-mass stellar models of \citet{Girardi00}
to those implemented in the Dartmouth stellar evolution database
\citep{Dotter08}, i.e. as large as $\sim$200\,K for solar metallicity. Note
however that the obtained differences in the magnitudes of these stars are much
less noticeable as is shown in the lower panel of Fig.~\ref{VLMS}. 
 
\begin{figure}
\includegraphics[angle=0,width=\columnwidth]{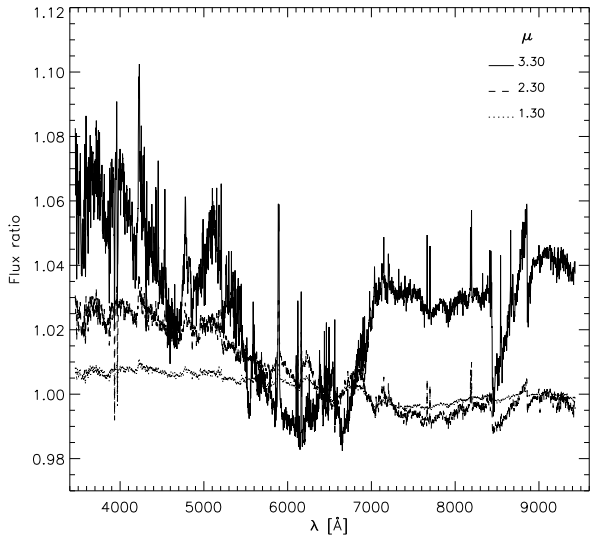}
\caption{Flux ratio between SSP SEDs, at their nominal resolution, of solar
metallicity and 10\,Gyr computed with a Unimodal IMF with slopes $\mu=$ 1.3,
2.3 and 3.3, adopting two different models for stars with $M<0.5$\,\Msol\
\citep{Girardi00, Pols95}. The spectra were normalized to unity at
$\lambda\lambda$ 6466-6480\,\AA. See the text for details.}
\label{SSP_VLMS}
\end{figure}

To explore the net effect on the synthesized SSP SEDs of adopting different
models for these low-mass stars we replaced the stars with $M<0.5$\,\Msol\ of
\citet{Girardi00} by those of \citet{Pols95}, i.e, the hottest by the coolest
models, respectively. All the stellar models were converted to the
observational plane following the same prescriptions described in
Section~\ref{sec:ingredients}. Fig.~\ref{SSP_VLMS} shows the flux ratio between
the SSP SEDs of solar metallicity and 10\,Gyr with Unimodal IMF of slope 1.3,
2.3 and 3.3, computed with these two sets of stellar models. For $\mu=$ 1.3,
i.e. the Salpeter case, we do not find any significant effect. For $\mu=$ 2.3
the effects are in general smaller than 3\%, with the largest impact seen for
the Na feature at $\sim$5800\,\AA.  Following the NaD Lick index definition for
this feature \citep{Worthey94}, the obtained index difference translates to an
observing error corresponding to a signal-to-noise around 30. Finally for
$\mu=$ 3.3 the effects in some features can be as large as $\sim$10\%, and for
that reason we do not consider safe such model SEDs.

\begin{table*}
\label{tab:SEDproperties}
\centering
\caption{Spectral properties and parameter coverage of the SSP SEDs}
\begin{tabular}{lc}
\hline 
\multicolumn{2}{c}{Spectral properties}\\
\hline                   
Spectral range     & $\lambda\lambda$ 3464.9 -- 9468.8\,\AA \\	
Spectral resolution& FWHM $=2.51$\,\AA, ($\sigma=64.1$\,km~s$^{-1}$ at 5000\,\AA, $\sigma=40.1$\,km~s$^{-1}$ at 8000\AA)\\	
Linear dispersion  & 0.9\,\AA/pix (54.0\,km~s$^{-1}$ at 5000\AA, 33.7\,km~s$^{-1}$ at 8000\AA)\\ 
Continuum shape    & Flux-scaled                    \\
Units &F$_{\lambda}$/L$_{\odot}$\AA$^{-1}$M$_{\odot}^{-1}$, L$_{\odot}=3.826\times10^{33}{\rm erg.s}^{-1}$ \\
\hline
\multicolumn{2}{c}{SSPs parameter coverage}\\
\hline
IMF type           & Unimodal, Bimodal, Kroupa universal, Kroupa revised\\
IMF slope (for unimodal and bimodal) & 0.3, 0.8, 1.0, 1.3, 1.5, 1.8, 2.0, 2.3, 2.8, 3.3 \\
Stellar mass range& 0.1 -- 100\,M$_{\odot}$ \\
Metallicity        & $-2.32$, $-1.71$, $-1.31$, $-0.71$, $-0.41$, $0.0$, $+0.22$ \\
%Age (\mbox{$\mbox{[Z/H]}=-2.32$})& $10.0 < t < 18$\,Gyr	(only for IMF slopes $\le$ 1.8)\\
%Age (\mbox{$\mbox{[Z/H]}=-1.31$})& $0.3 < t <13$\,Gyr; $\mu \leq 1.3$ (SAFE range for non MILES/CaT spectral ranges)	   \\
Age & $0.06 < t < 18$\,Gyr \\ 
SAFE SSP SEDs for non MILES/CaT spectral ranges & (\mbox{$-0.71\le\mbox{[Z/H]}\le+0.22$} \& $0.06 < t < 18$\,Gyr \& $\mu \leq $2.3) \\
\hline
\end{tabular}
\end{table*}

\begin{figure}
\includegraphics[angle=0,width=\columnwidth]{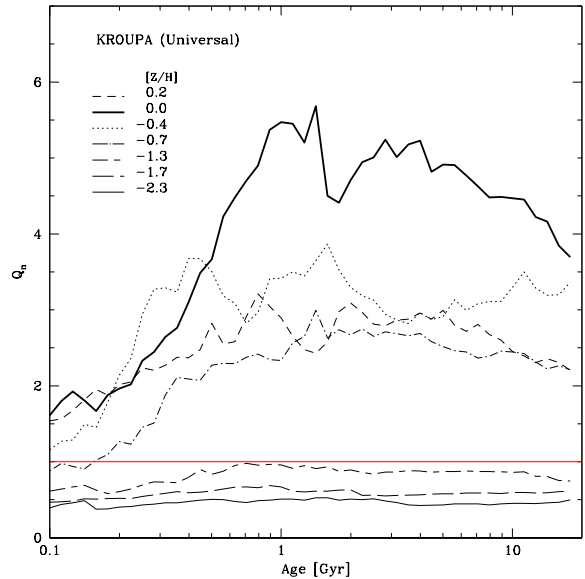}
\caption{The quality parameter, Q$_n$, defined in \citet{Vazdekis10} as a
function of the SSP age (in Gyr) for different metallicities, indicated by the
line types as quoted in the inset. For computing the SEDs we use a Kroupa
Universal IMF. SSPs with Q$_n$ values larger than 1 can be safely used (see the
text for details).}
\label{fig:Qn}
\end{figure}

To estimate the reliability of the newly computed SEDs and the SSP parameter
regions where these models can be considered safe, we use the normalized quality
parameter $Q_n$ defined in \citet{Vazdekis10}. To compute $Q_n$ for each SSP we
make use of our algorithm for synthesizing a representative stellar spectrum for
a given set of atmospheric parameters. The larger the total number of stars and
the shorter the distance of their parameters to the requested ones for
each star along the isochrone, the larger the quality parameter of the
synthesized SSP spectrum is. The obtained value is normalized with respect to a
minimum acceptable value, which comes from a poor, but still acceptable,
parameter coverage of the stellar library. We refer the interested reader to 
\citet{Vazdekis10} for a full description of the method. $Q_n$ allows us to
obtain a quality measure of the SSP SED, due to the atmospheric parameter
coverage of the stellar library feeding the models. 

Figure~{\ref{fig:Qn}} shows the value of $Q_n$ as a function of the SSP age (in
Gyr) for different metallicities (different line types) and adopting a Kroupa
Universal IMF. SSP SEDs with $Q_n$ values above 1 can be considered of
sufficient quality, and therefore they can be safely used. Note that these
values only apply to the spectral ranges not covered by MILES and CaT SEDs,
which in general provide larger values. As expected, the higher $Q_n$ values
are obtained for solar metallicity. $Q_n$ reaches a value of $\sim$5 for
stellar populations in the range 1 -- 10\,Gyr.  Outside this age range the
quality decreases as low-mass and hotter MS stars are less numerous in the
MIUSCAT database (see Fig.\,\ref{miusc}). Both, for the supersolar and for the
low metallicities we obtain lower, but acceptable, values. 
For \Zh=-1.31 we obtain $Q_n$ values that are  slightly below 1,  but
should not be considered safe, particularly for detailed spectroscopic studies
that employ line-strengths or full spectrum-fitting approach. Finally, for
lower metallicities the MIUSCAT SEDs are not reliable, though these models
might still be used for some applications such as, e.g., broad-band colours (see
Section~\ref{sec:behaviour}).
Note that in general the $Q_n$ values obtained for the MIUSCAT SEDs are lower than those of MILES SEDs (see
second panel of Fig.~6 in \citealt{Vazdekis10}), for which we obtain acceptable
$Q_n$ values for metallicities as low as \Zh = -2.3.
Table~\ref{tab:SEDproperties} summarizes the safe ranges for the MIUSCAT SED
spectral ranges that are not covered by MILES and CaT SEDs. For the latter we
refer the reader to \citet{Vazdekis10} and the MILES webpage.

%%%%%%%%%%%%%%%%%%%%%%%%%%%%%%%%%%%%%%%%%%%%%%%%%%%%%%%%%%%%%%%%%%%%%%%%%%%%%%%%

\section{Behaviour of the models}
\label{sec:behaviour}

The behaviour of our model SEDs as a function of the SSP parameters in the CaT
and MILES wavelength ranges have been shown in \citet{Vazdekis03} and
\citet{Vazdekis10}. We refer the reader to these papers for an extensive
analysis. In this section we focus on the new additions from expanding the
spectral coverage of the model SEDs. Among these new features we discuss here
our colour measurements, which are obtained from convolving the MIUSCAT SSP SEDs
with the corresponding filter responses. 

\begin{figure*}
\includegraphics[angle=0,width=\columnwidth]{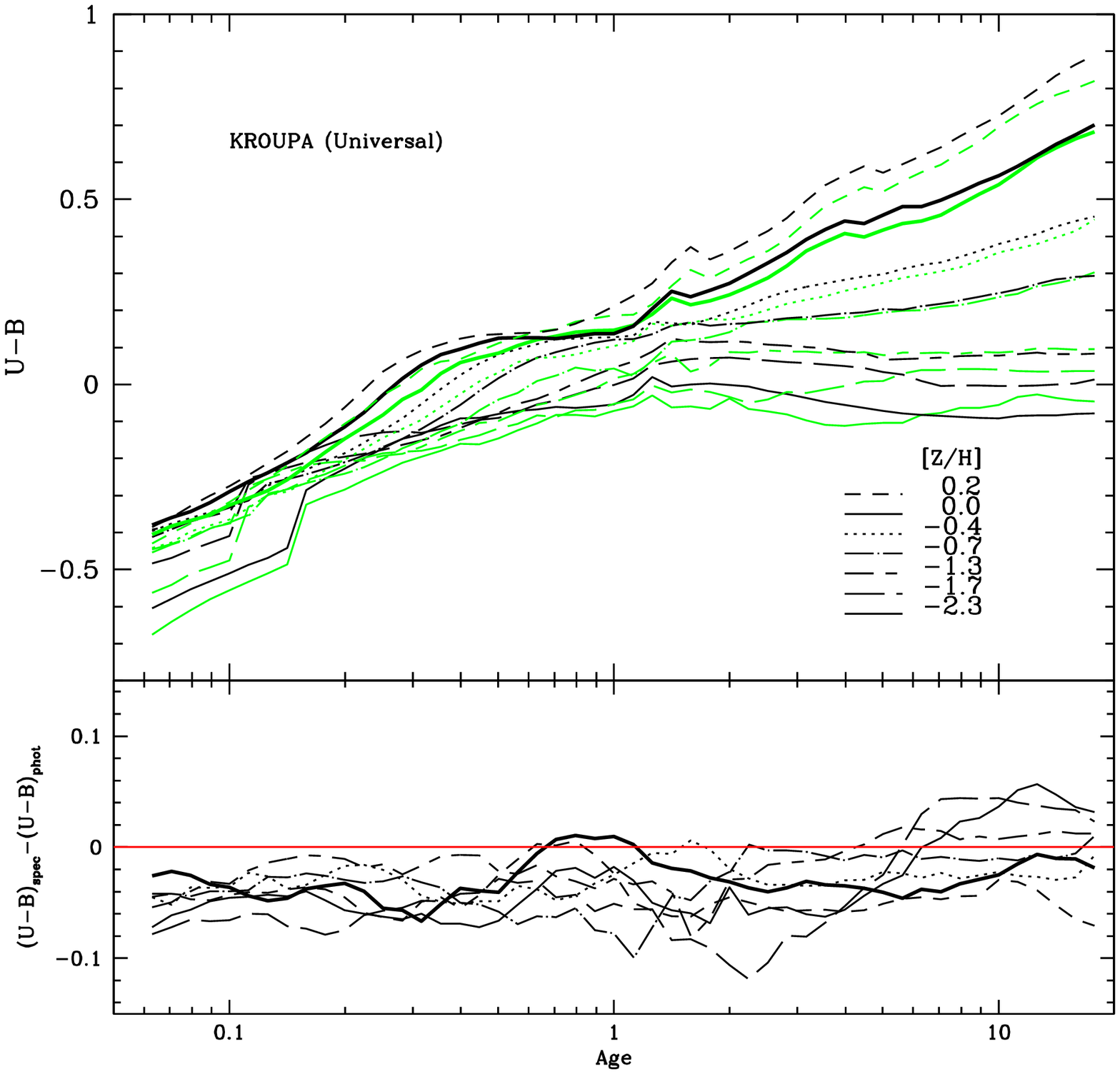}
\includegraphics[angle=0,width=\columnwidth]{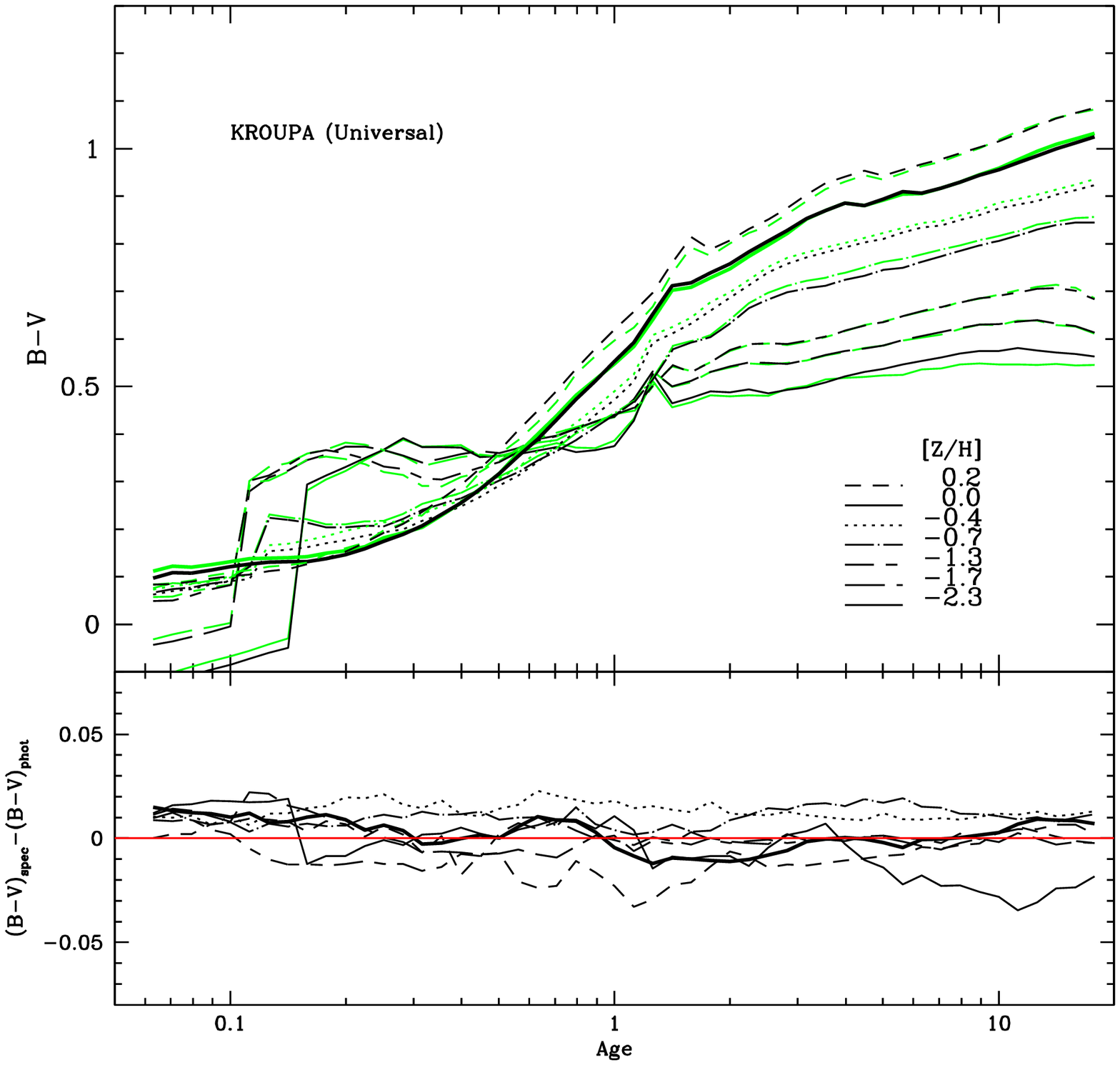}
\includegraphics[angle=0,width=\columnwidth]{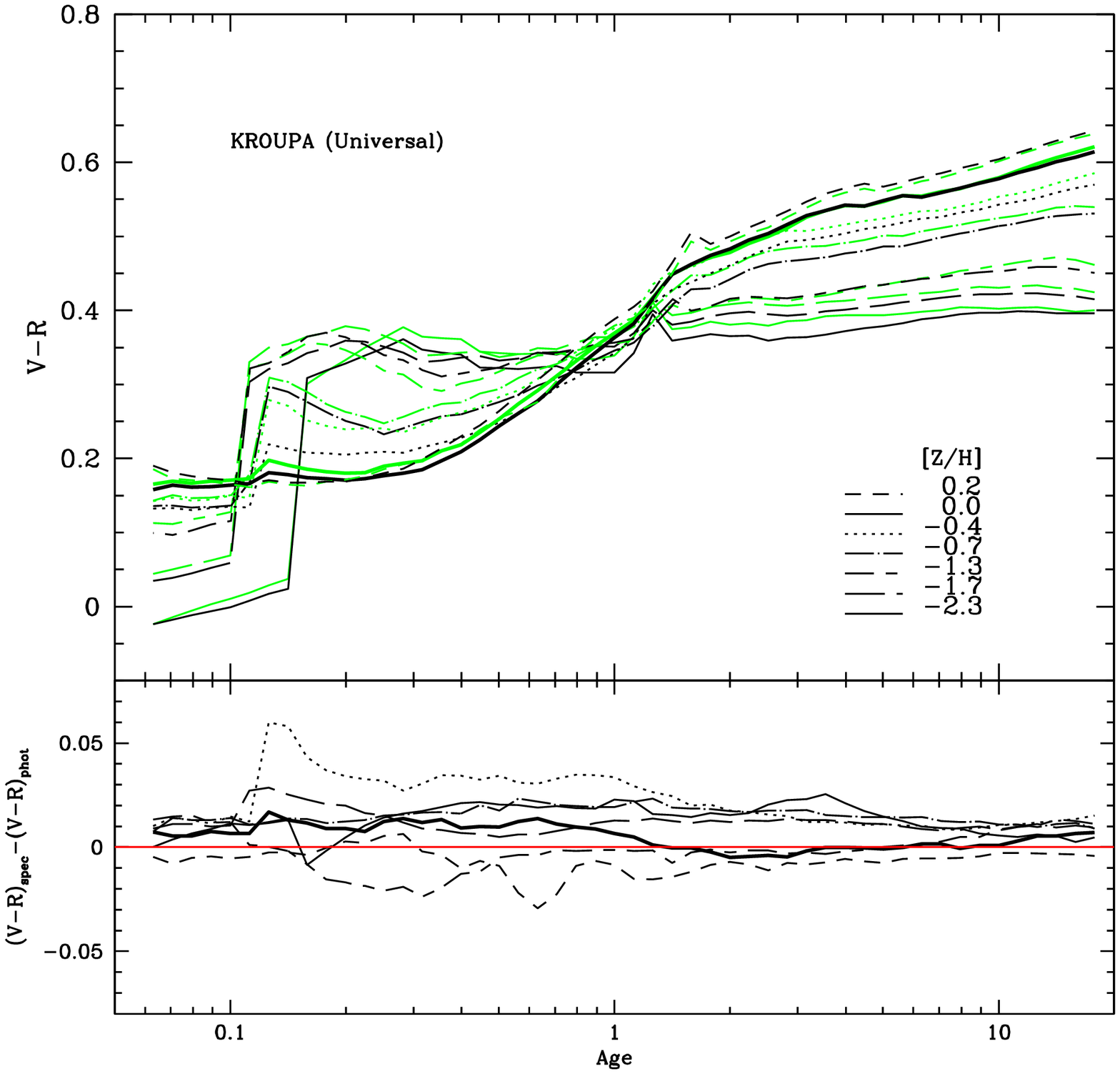}
\includegraphics[angle=0,width=\columnwidth]{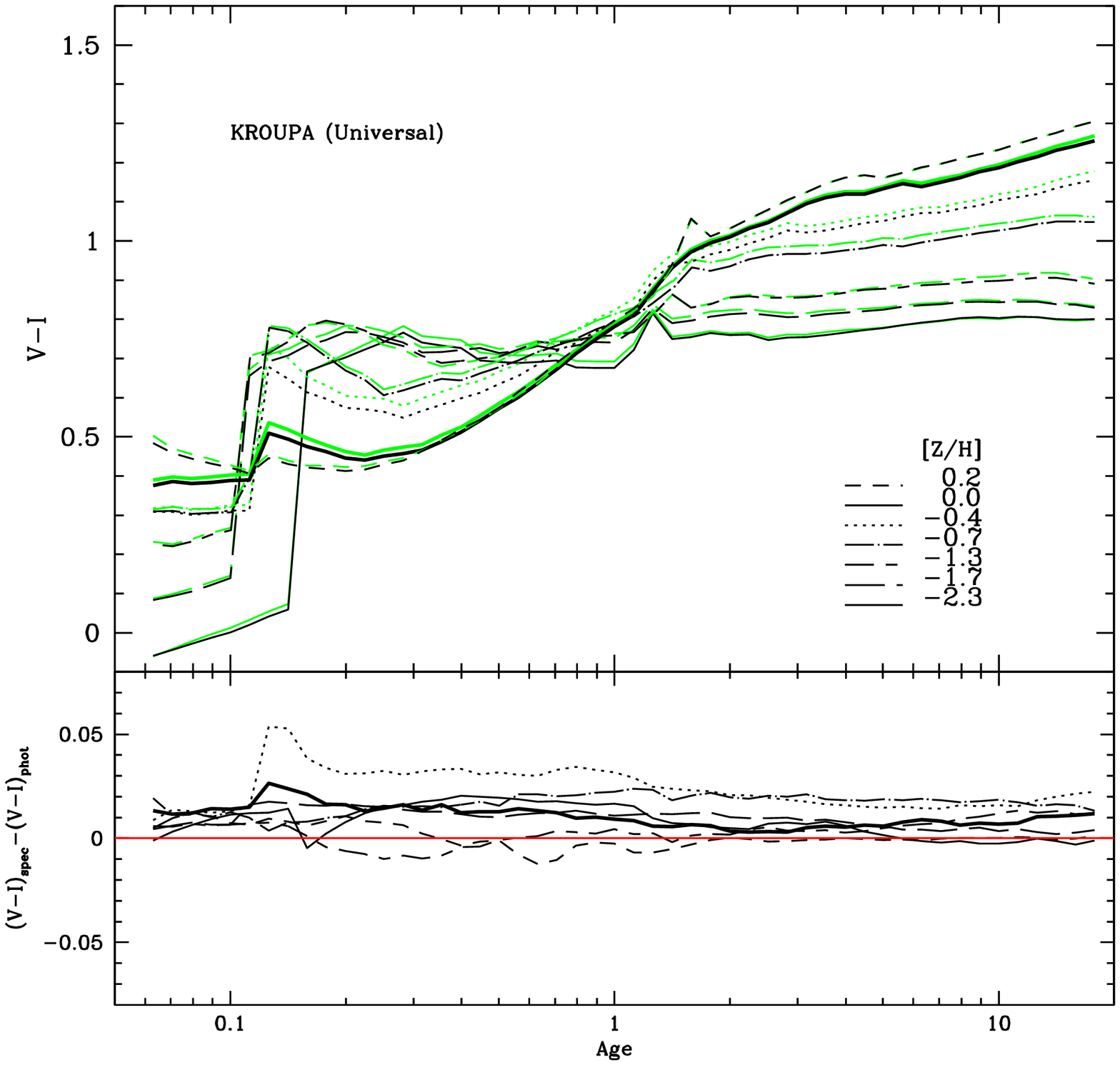}
\caption{We plot the synthetic broad band $U-B$ $B-V$ $V-R$ and $V-I$
  colours (black lines) 
derived from the MIUSCAT SSP SEDs for different ages and metallicities (as
indicated within the panels) and the Kroupa Universal IMF with the zero point
set with \citet{Hayes85} Vega spectrum. The green lines represent the photometric
colours computed by \citet{Vazdekis96}, as updated in \citet{Vazdekis10},
which are calculated on the basis of the empirical photometric relations described in
Section~\ref{sec:ingredients}. The obtained residuals are also shown for each
colour.}
\label{SSPcolors}
\end{figure*}

Fig.~\ref{SSPcolors} shows the broad-band Johnson-Cousins
\citep{Buser78, Cousins80} colours, $U-B$, $B-V$,
$V-R$ and $V-I$, derived from the MIUSCAT SSP SEDs for different ages and
metallicities, adopting a Kroupa Universal IMF with the zero point set with the
Vega spectrum of \citet{Hayes85}. For comparison we also show the photometric
colours of \citet{Vazdekis96}, as updated in \citet{Vazdekis10}, which are
calculated on the basis of the relations described in
Section~\ref{sec:ingredients}, which come from extensive empirical photometric
libraries. In this figure $B-V$ is the only colour that can be fully measured
within the MILES spectral range, with no need to use the MIUSCAT wavelength range
extension.

In general we see that the obtained residuals are as small as typical zero-point
uncertainties ($\sim$0.02 mag) for nearly all the metallicities for the $B-V$,
$V-R$ and $V-I$ colours. For the $B-V$ colour this can be attributed to the very
high flux-calibration quality of the MILES stellar spectra as was shown in
\citet{Sanchez06}. On the other hand the good agreement obtained for the $V-R$
and $V-I$ colours should be attributed to our method for scaling and joining the
different spectral ranges. Finally the largest residuals are obtained for the
$U-B$ colour. Note however that the measurement of the $U$ magnitude is very
sensitive to the various filter definitions for this band as is shown in
Paper~II. Our photometric $U-B$ predictions are mainly obtained with the aid of
the relations of \citet{Alonso96, Alonso99}, as for the other filters. However
these $U-B$ relations are not completely homogeneous with the relations obtained
by these authors for the other filters. We refer the interested reader to these
authors for a detailed description of the methods to derive these relations.

\begin{figure}
\includegraphics[angle=0,width=\columnwidth]{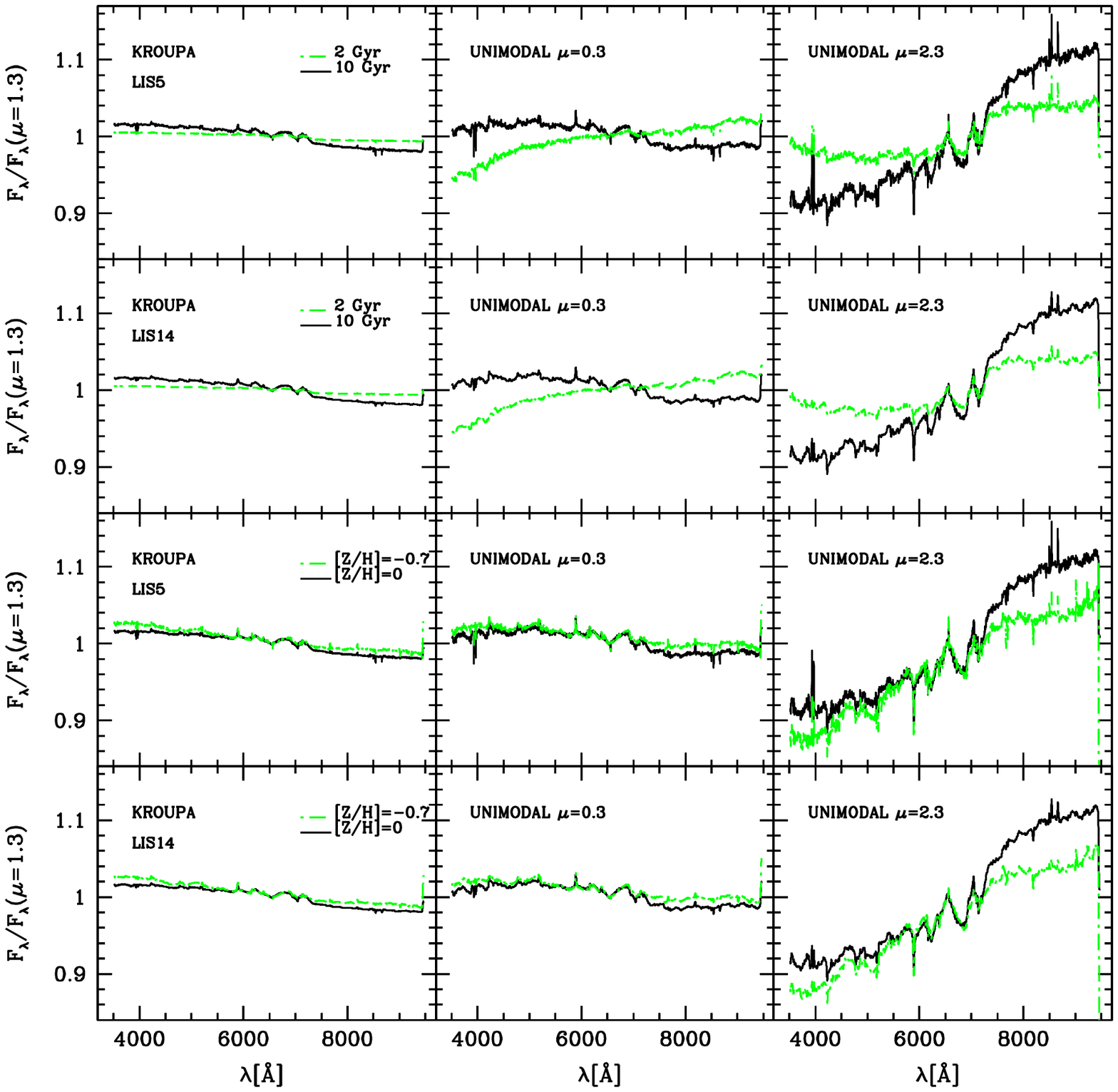}
\caption{ Flux ratio between SSP models with various IMFs (Kroupa Universal in
the left panels, Unimodal with slope $\mu=0.3$ in the center and Unimodal with
slope $\mu=2.3$ in the right) with the corresponding model with Unimodal IMF
and $\mu=1.3$ (i.e. Salpeter). The first row of panels shows models with solar
metallicity and two ages (2 and 10\,Gyr), as indicated in the legend of the
left panel. All the models were smoothed to match the resolution of the
\citet{Vazdekis10} LIS-5.0\AA\ line-index system, i.e. FWHM$=$5.0\AA. The
spectra were normalized to unity at $\lambda\lambda$ $\sim$6466-6480\,\AA. We
keep the same scale for all the flux ratios. The second row of panels shows the
same flux ratios but with the SSP SEDs smoothed to match the spectral
resolution of the LIS-14.0\AA\ system of \citet{Vazdekis10}, i.e.
FWHM$=$14\,\AA. Finally, the third and fourth row of panels show the flux
ratios of SSP SEDs with 10\,Gyr and two metallicities (\Zh$=$0.0 and
\Zh$=$-0.7), as indicated in the legend of the first panel, with all the models
smoothed to match the LIS-5.0\AA\ and LIS-14.0\AA\ systems, respectively.}
\label{fratio}
\end{figure}

Another interesting aspect that can be studied with the newly synthesized
MIUSCAT SSP SEDs is the variation of the colours and spectral features with the
IMF, due to the expanded spectral range. The first row of panels of
Fig.~\ref{fratio} shows the  ratio obtained by dividing the SSP SEDs of solar
metallicity and various IMFs with respect to the SED computed with a Unimodal
IMF of slope 1.3 (Salpeter case), for two different ages (2 and 10\,Gyr). All
the spectra are smoothed to 5\,\AA\ (FWHM), i.e. the constant resolution that
characterizes the LIS-5.0\AA\ system of index measurements introduced in
\citet{Vazdekis10}. The first panel shows that the models constructed with a
Kroupa Universal IMF is rather similar to the Salpeter case for the two ages.
The larger, though small, difference is only seen for redder
wavelengths  
 for the 10\,Gyr old model. This colour trend is
further accentuated for the IMF with the largest slope ($\mu=2.3$), i.e., the
larger the slope the larger the effect is. Also, the effect is more pronounced
as the age increases.

This effect is originated by the large contribution  of low-mass MS stars in
the steep IMF models. Being their SEDs red, the resulting total flux of the
model is redder than models with flatter IMF slopes. Unlike the blue colours,
the redder ones are clearly sensitive to the IMF. A similar result was shown in
\citet{Vazdekis96}. 

Apart of the colours we also see in the bottom panels a significant
IMF-sensitivity for various absorption spectral features and molecular bands.
Among the most prominent features we identify the Ca\,H\,\&\,K around
$\sim$3950\,\AA, Ca{\sc I} at $\sim$4227\,\AA, the Na{\sc I} doublets at
$\sim$5900\,\AA\ and $\sim$8200\,\AA, H$_{\alpha}$ and the Ca{\sc II} triplet
around $\sim$8600\,\AA. The IMF-sensitivity of all these features, and other
features in redder wavelengths, have been recently discussed in \citet{Conroy11}
(see references therein). We also find emphasized IMF-sensitivities in the TiO
molecular bands redwards 5900\,\AA. These include the prominent features seen at
$\sim$5950\,\AA, $\sim$6200\,\AA, $\sim$6600\,\AA, $\sim$7200\,\AA\ and
$\sim$8900\,\AA. The first two TiO bands are included in the Lick system of
indices (i.e., TiO$_1$ and TiO$_2$) (Worthey94). Indices for the other TiO bands
have been discussed and modelled for stellar populations in e.g.,
\citet{Schiavon97}, \citet{Faber80} and \citet{Vazdekis03}. Moreover the slope
of the SSP SED around the Ca{\sc II} triplet at $\sim$8600\,\AA\ has also been
shown to be sensitive to the IMF slope \citep{Vazdekis03,Cenarro09}. In the
latter paper this slope has been modelled by defining a specific index, named
sTiO. An index-index diagram, which includes this index and the Ca{\sc II}
triplet at $\sim$8600\,\AA, has been proposed in \citet{Cenarro03} to diagnose
the IMF slope.  

It is worth noting that whereas the Ca{\sc II} triplet decreases with increasing
IMF slope, the nearby Na{\sc I} doublet at $\sim$8200\,\AA\ increases. Therefore
the study of these two features constitutes a powerful diagnostic for
constraining the IMF. The latter feature, which falls in the spectral region not
covered by the MILES and CaT models, represents a very important improvement of
the newly synthesized MIUSCAT SSP SEDs. We focus on this aspect in Section
~\ref{sec:index}. The sensitivity of the Ca{\sc II} triplet to the IMF in SSP
models was extensively shown and discussed in \citet{Vazdekis03}. See also
\citet{Schiavon00} and \citet{Conroy11}. The IMF-sensitivity of the Na{\sc I}
doublet at $\sim$5900\,\AA\ was also shown in \citet{Vazdekis96},
\citet{Schiavon00} and \citet{Conroy11}. 

As expected, the flux ratio features that are seen at 5\,\AA\ resolution are
generally less pronounced when the spectra are smoothed to match the lower
resolution of the LIS-14.0\AA\ system, i.e. FWHM$=$14\,\AA, as shown in
the second row of panels of  Fig.~\ref{fratio}. This extent shows that the
resolution of the data or the velocity dispersion of the galaxies should be
taken into account when studying the IMF effects.

In the third and fourth row of panels of Fig.~\ref{fratio}
we fix the age of the SSP SEDs to 10\,Gyr in order to study the IMF sensitivity
as a function of the metallicity at the resolutions of the LIS-5.0\AA\ and
LIS-14.0\AA\ systems, respectively. Overall we see that most of the flux ratio
features tend to be less pronounced in the metal-poor model. Interestingly,
this is not the case for the Na{\sc I} doublet at $\sim$8200\,\AA.

\begin{figure}
\includegraphics[angle=0,width=\columnwidth]{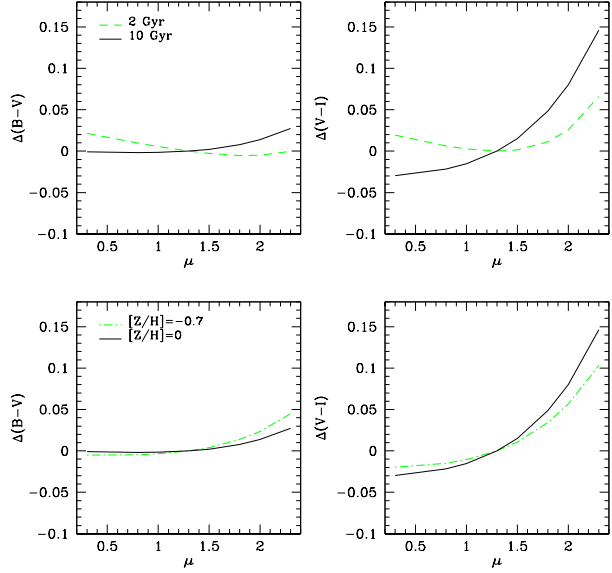}
\caption{$B-V$ (left panels) and $V-I$ (right panels) colour difference with
respect to the SSP models with Unimodal IMF with $\mu=1.3$ (i.e. Salpeter), as
a function of the IMF slope. In the upper panels two ages (2 and 10\,Gyr) are
considered, whereas in the lower panels we show the results for two
metallicities (\Zh$=$-0.7 and 0).} \label{deltacolBVI} \end{figure}

\begin{figure}
\includegraphics[angle=0,width=\columnwidth]{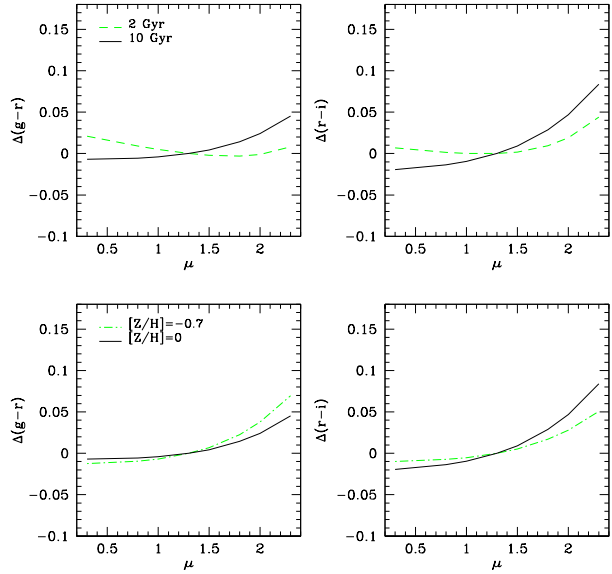}
\caption{Same as in Fig.~\ref{deltacolBVI} but for the $g-r$ (left panels)
and $r-i$ (right panels) colours of the SDSS system.}
\label{deltacolgri}
\end{figure}

In Fig.~\ref{deltacolBVI} we show the difference in $B-V$ and $V-I$ colours
with respect to the corresponding values for the models with Unimodal IMF with
$\mu=1.3$ as a function of the IMF slope, for the two ages and two
metallicities. As it can be inferred from  Fig.~\ref{fratio}, which shows
the larger IMF-sensitivity of the redder spectral regions, the $V-I$ colour
differences are significantly larger than those obtained for $B-V$. We also see
that the main increase of the colour difference takes place for IMFs with
$\mu>2$. Finally, a similar behaviour is found for the colours of the SDSS
filters as shown in Fig.~\ref{deltacolgri}. 

%%%%%%%%%%%%%%%%%%%%%%%%%%%%%%%%%%%%%%%%%%%%%%%%%%%%%%%%%%%%%%%%%%%%%%%%%%%%%%%%

\section{Applications}
\label{sec:applications}

In this section we show several applications to illustrate the potential use of
MIUSCAT SSP SEDs. We only focus on those colours and features that we were not
able to measure with the MILES/CaT SEDs but are now possible with the present
extension of the spectral coverage of the models. These examples include
broad-band colours, a newly-defined line-strength spectroscopic absorption index
and survey-oriented spectrophotometric applications.

\subsection{Galactic globular cluster colours}
\label{sec:MW}

\begin{figure*}
\includegraphics[angle=0,width=\columnwidth]{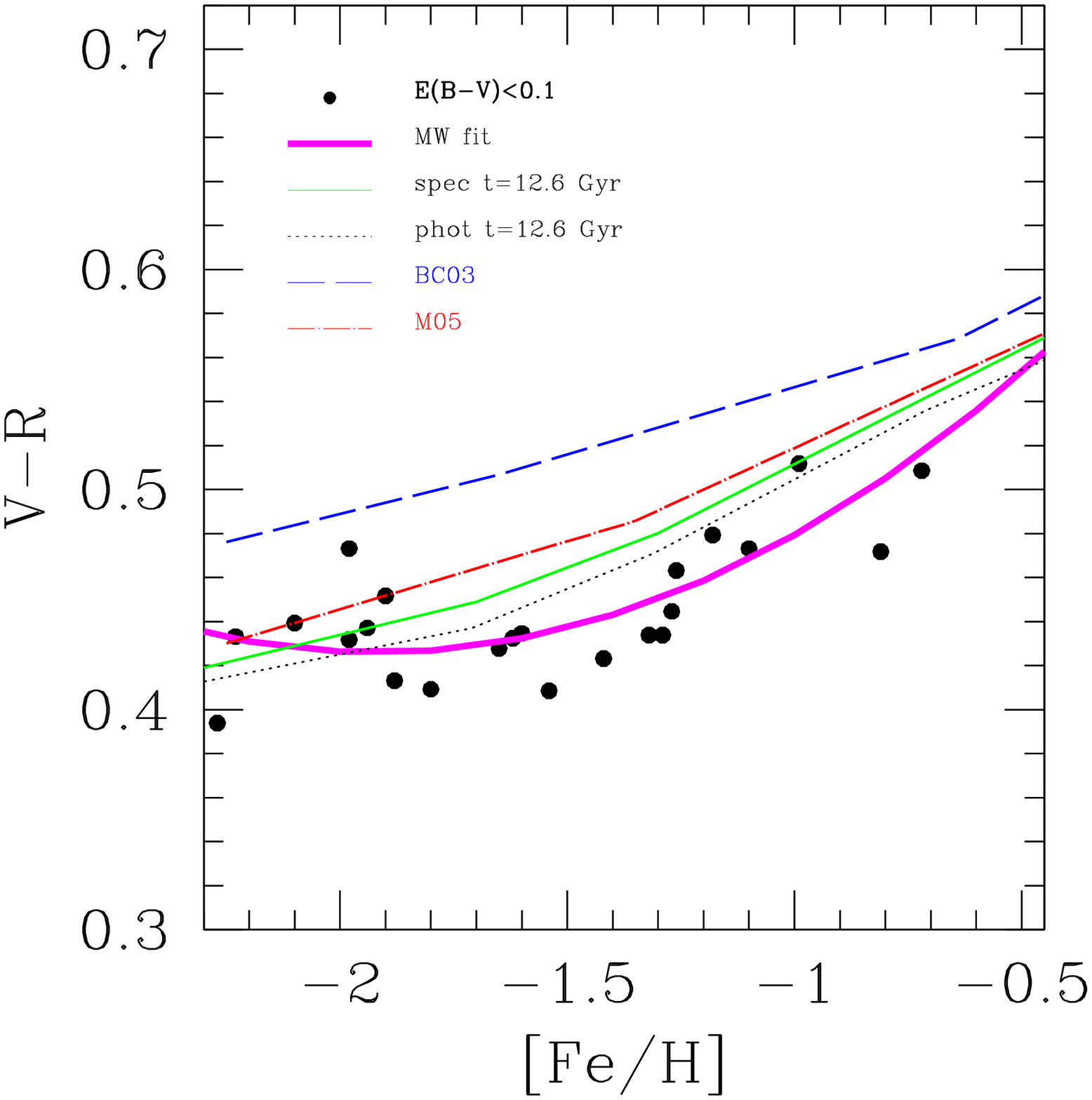}
\includegraphics[angle=0,width=\columnwidth]{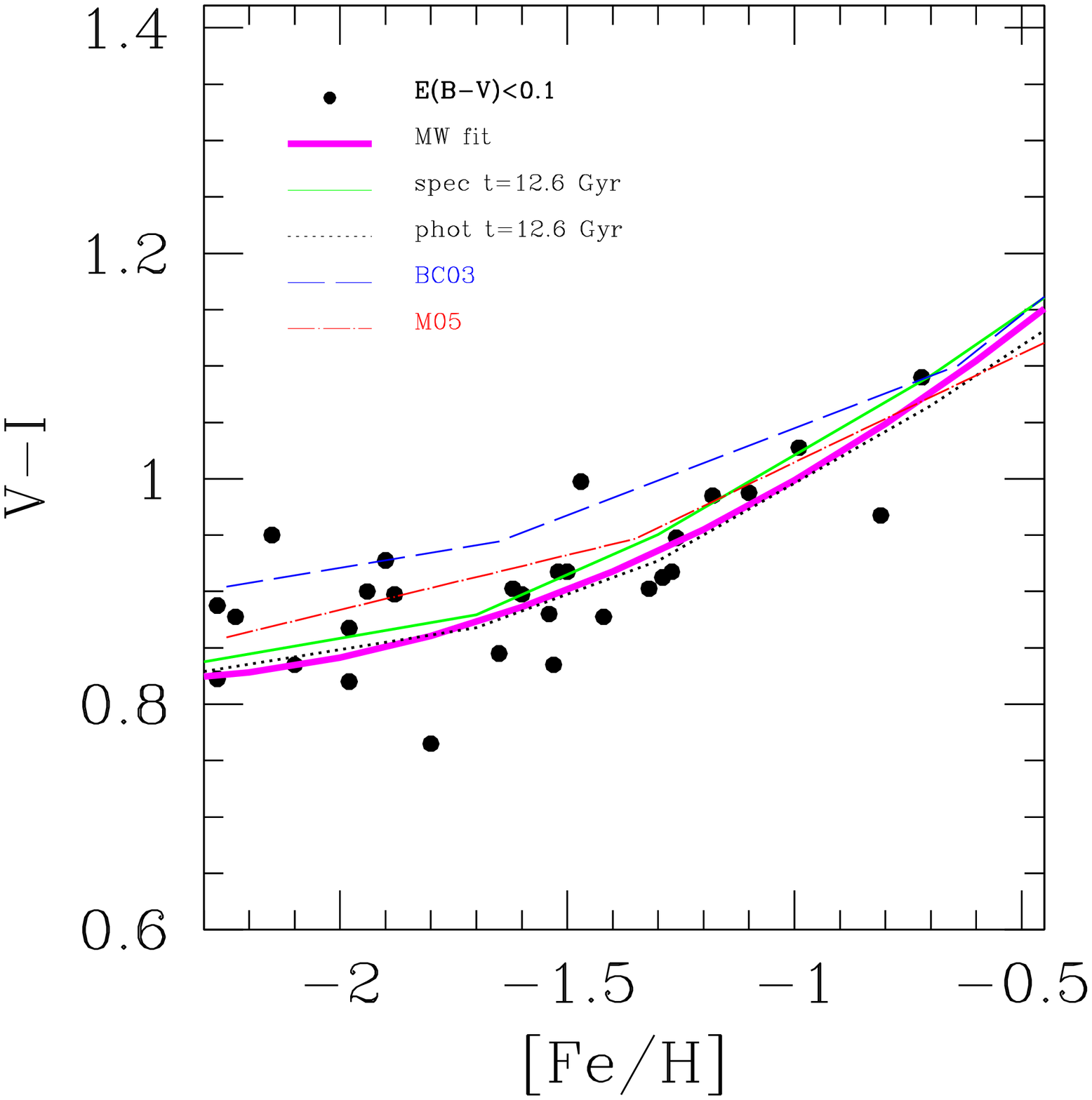}
\caption{Comparison of $V-R$ (left panel) and $V-I$ (right panel) colours from
models of $\sim$12\,Gyr and MW globular cluster data from the \citet{Harris96}
catalog. We only plot clusters with small extinction values ($E(B-V) < 0.1$).
The colours are plotted against the  metallicity from the same
catalog. The thick magenta line represents an order 2 polynomial fit to the
data. All the models are computed adopting a Salpeter IMF. In each panel the
dotted black line represents the photometric colour prediction of
\citet{Vazdekis96}, as updated in \citet{Vazdekis10}, whereas the thick solid
green line represents the synthetic colour derived from the MIUSCAT SEDs. We
also show for comparison the \citet{BC03} and \citet{Maraston05} models in
long-dashed blue and dot-dashed red lines, respectively.}
\label{MW_VRI}
\end{figure*}

Fig.~\ref{MW_VRI} shows a comparison of the $V-R$ and $V-I$ colours with the
Milky Way globular cluster data with $E(B-V) < 0.1$, from the catalog of
\citet{Harris96}. The colours are plotted against  metallicities from
the same catalog. All the models have $\sim$12\,Gyr and adopt a Salpeter IMF. We
provide two flavours of models: the photometric predictions of
\citet{Vazdekis96}, as updated in \citet{Vazdekis10}, and the synthetic colours
derived from the MIUSCAT SEDs. We see that these two sets of predictions agree
very well each other, matching the observations for the expected ages and
metallicities according to independent measurements from CMD studies.  We also
show for comparison the colours from the models of \citet{BC03} and
\citet{Maraston05}. The latter provide too red values in comparison to our
predictions and the data. \citet{Maraston11} have shown that models employing
theoretical stellar libraries tend to provide such redder colours in comparison
to those based on empirical libraries, as in the present work. 

Further colour comparisons for globular clusters in the Galaxy and in M\,31 are
shown in Paper~II. Both colours in the Johnson-Cousins and in the SDSS systems
are discussed. We also show in that paper an extensive and comprehensive study
for the early-type galaxies of the SDSS out to $z\sim0.5$.  

\subsection{Line-strength index definitions}
\label{sec:index}

We illustrate in this section the potential use of the MIUSCAT SSP SEDs for
defining new line-strength indices. For this purpose we focus on the most
interesting feature that can be found in the spectral region that was not
covered by the MILES and CaT SEDs, i.e., the Na{\sc I} doublet at
$\sim$8200\,\AA\ (see Section~\ref{sec:behaviour}). This feature, together with
the FeH Wing-Ford band absorption at 9916\,\AA, has been recently used to
suggest a dwarf-enriched scenario for massive cluster elliptical galaxies
\citep{vanDokkum10}. There is also evidence of Na{\sc I} gradients suggesting
concentrations of metal-rich dwarfs toward the centers of giant
galaxies \citep{Carter86}.

\begin{figure}
\includegraphics[angle=270,width=\columnwidth]{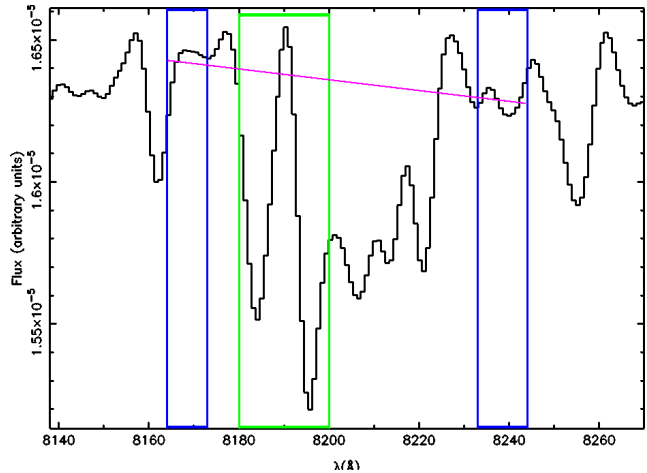}
\caption{New index definition for the Na{\sc I} doublet around 8200\,\AA\
(Na{\sc I}8200A). The feature bandpass is indicated in green. The continuum flux level
(dashed magenta line) is defined by the pseudocontinua at either side of the
feature (indicated in blue). The plotted SED corresponds to a MIUSCAT SSP model
of solar metallicity, 10\,Gyr and Unimodal IMF with slope 1.3 (i.e. Salpeter).
The spectrum has been smoothed to match the spectral resolution of the
LIS-5.0\AA\ line-index system, i.e., 5\,\AA\ (FWHM).} 
\label{NaI} 
\end{figure}

We follow here the same scheeme used to define the Lick-style indices of the
optical spectral range \citep{Worthey94}. We use a feature bandpass and two
pseudocontinua bandpasses, which are located at either side of the feature, in
order to establish the flux level of the continuum at the feature. These
bandpasses are illustrated in Fig.~\ref{NaI}, whereas the limiting wavelengths
are listed in Table~\ref{tab:NaI8200Adefinition}. This is a modified version of
the index proposed for this feature by \citet{Schiavon97} that avoids the TiO
bandhead absorption at 8205\,\AA, which becomes increasingly stronger than the
Na{\sc I} lines for decreasing temperatures in giants (see also
\citealt{Carter86}). Our index represents an improvement over previous
definitions in this respect (e.g., \citealt{Faber80}, \citealt{Serven05};
\citealt{Conroy11}). In fact \citet{Alloin89} propose an alternative discussion
about dwarf/giants star content in the semi-stellar nucleus of M\,31, which is
based on the metallicity, precisely due to the TiO absorption contamination.
Note however that our index definition is certainly an advantage when
fitting low velocity dispersion galaxies as this contamination cannot be
avoided in more massive systems. Another advantage of the Na{\sc I}8200A index
is that we adopt a significantly wider pseudocontinua ($\sim$10\,\AA) than in
the \citet{Schiavon97} index ($<$1\,\AA), which makes it more robust against
the signal-to-noise of the observational data.

\begin{figure}
\includegraphics[angle=0,width=\columnwidth]{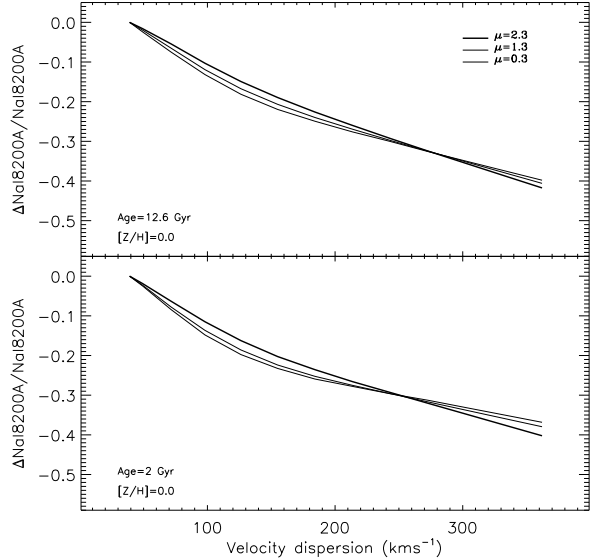}
\caption{Broadening correction for the newly defined Na{\sc I}8200A index,
$\Delta$Na{\sc I}8200A/Na{\sc I}8200A, in SSPs with Unimodal IMF and varying slopes. The
solid lines become thicker with increasing slope.The upper panel shows the results for a 12.6\,Gyr model, whereas in
the lower panel we show the results for a 2\,Gyr, the two for solar metallicity.
We have broadened the model spectra by convolving with gaussians from the
nominal resolution, i.e.\,$\sigma$=39\,\kms (FWHM=2.51\,\AA) at 8200\,\AA, up to
$\sigma$=360\,\kms, in steps of 30\,\kms.}
\label{fig:nad_sigma}
\end{figure}

It is worth noting that we do not intend here to provide an optimal index
definition that requires a specifically dedicated paper, which is out of the
scope of this work. A good example can be found in \citet{Cervantes09}, where we
proposed a new index for the H$\beta$ feature, with an unprecedented ability to
disentangle the age of the stellar populations. Nonetheless, we have tried
several index definitions for this feature, which turned out to be extremely
sensitive to the velocity dispersion smearing of the two absorption lines of
this doublet. In fact these lines blend in a single feature for velocity
dispersions around 200\,\kms. Therefore it is important to assess its dependence
on this effect if this feature were to be used for constraining the IMF. Our
experiments show that the index definition is more robust against such smearing
if these two lines are included within the feature bandpass. Note also that our
feature bandpass is well isolated from the two selected pseudocontinua
bandpasses. Fig.~\ref{fig:nad_sigma} shows the behaviour of the Na{\sc I}8200A index
against the velocity dispersion for models with Unimodal IMF and varying slope.
The plots show the results for two different ages and solar metallicity, but
similar patterns are seen for other values of these two SSP parameters. As
expected, we see that the index decreases its strength with decreasing
resolution, but this dependence does not vary significantly with the IMF slope.

\begin{table}
\centering{
\caption{Bandpass limiting wavelengths for the Na{\sc I}8200A index.}
\label{tab:NaI8200Adefinition}
\begin{tabular}{@{}cc@{}}
\hline                    
Feature   & Pseudo-continuum \\                     
bandpasses (\AA)   & bandpasses (\AA) \\ 
\hline                     
8180--8200 & 8164--8173   \\       
           & 8233--8244   \\ 
%8180--8186 & 8164--8173   \\       
%8192--8198 & 8233--8244   \\ 
\hline
\end{tabular}
}
\end{table}

\begin{figure}
\includegraphics[angle=0,width=\columnwidth]{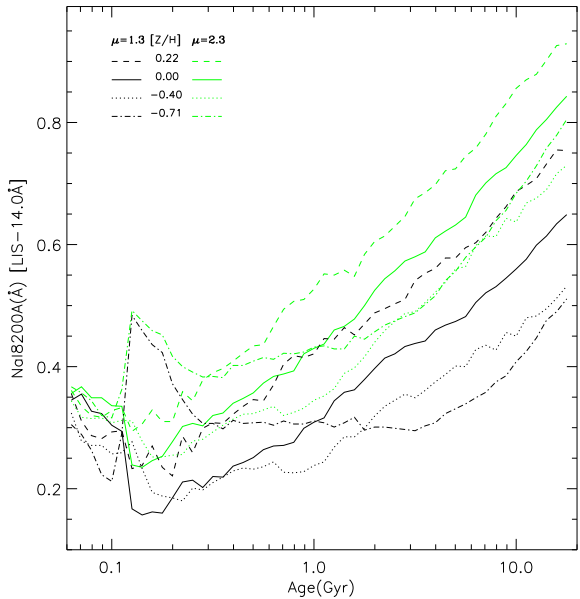}
\caption{Behaviour of the Na{\sc I}8200A index as a function of age for SSP models of
different metallicities, as indicated in the legend, for a Unimodal IMF with
slope $\mu=1.3$ (Salpeter) and $\mu=2.3$, plotted in black and green lines,
respectively. All the index measurements are performed in the LIS-14.0\AA\
system (FWHM=14\,\AA).}
\label{fig:nadb}
\end{figure}

Although not shown here, we also find that the Na{\sc I}8200A index is
extremely robust against uncertainties in the spectrum shape, which might
result from, e.g., flux-calibration issues. The differences in strengths
obtained from measuring the index in both an SSP SED with solar metallicity and
13\,Gyr with flux-calibrated response and the same model SED after removing its
continuum is virtually nil. This is expected as the index definition
encompasses only 80\,\AA, i.e. from the bluest wavelength of the blue
pseudocontinuum to the reddest wavelength of the red pseudocontinuum.

Fig.~\ref{fig:nadb} shows the behaviour of the Na{\sc I}8200A index as a
function of age and metallicity for SSP SEDs with Unimodal IMF with two
different slopes, $\mu=1.3$, i.e. Salpeter, and steeper, $\mu=2.3$. We use for
these measurements the LIS-14.0\AA\ system defined in \citet{Vazdekis10}. This
system has a flux-calibrated response curve and a constant resolution,
FWHM$=$14\,\AA, along the whole spectral range. At $\sim$8200\,\AA, where this
feature is located, this resolution translates to $\sigma$$\sim$220\,\kms,
which is appropriate for massive galaxies.  Although the Na{\sc I}8200A index
depends on the metallicity and age, it is particularly sensitive to the IMF
slope,  as expected from Fig.~\ref{fratio}. These results are in good
agreement with \citet{Schiavon00}. Fig.~\ref{fig:nadb} also shows that the
sensitivity to the IMF increases with increasing age and decreasing
metallicity. According to the formalism of \citet{Cardiel98} we obtain for this index an
error of $\sim$0.2 with a S/N(\AA$^{-1})$$\sim$30, for an old and solar
metallicity model. Such error is similar to the index difference obtained when
varying the IMF slope from 1.3 to 2.3 (see Fig.~\ref{fig:nadb}). This shows
that to apply this index for constraining the IMF in massive galaxies we require spectra with
significantly higher S/N. Note however that such IMF sensitivity might be
masked by similar index variations due to {\mbox{$\mbox{[Na/Fe]}$}} abundance
ratio, as shown in \citet{Conroy11}. These authors studied the sensitivity of
this feature to both {\mbox{$\mbox{[Na/Fe]}$}} and
{\mbox{$\mbox{[$\alpha$/Fe]}$}} abundances in SSP models of 13\,Gyr.

\begin{figure} 
\includegraphics[angle=0,width=\columnwidth]{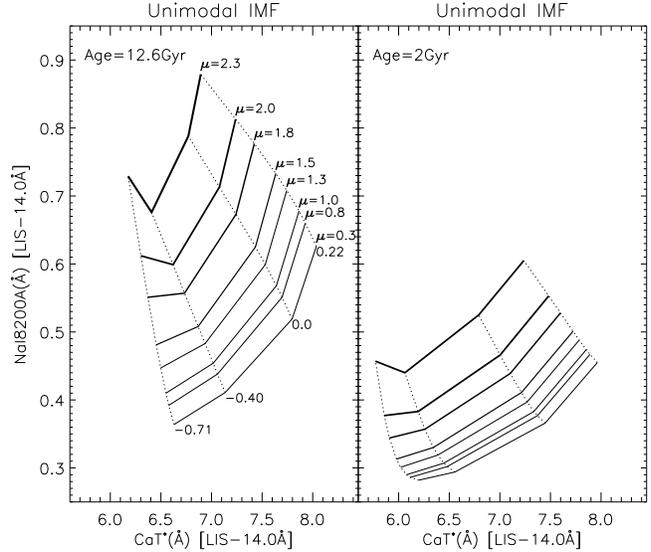}
\caption{Na{\sc I}8200A vs. CaT$^{*}$ diagnostic diagram to constrain the IMF slope.
We adopt a Unimodal IMF for all the models. The CaT$^{*}$ index is defined in
\citet{Cenarro01}. The solid lines indicate models with
constant IMF slope, the thicker the line the higher the slope
($\mu=$0.3,0.8,1.0,1.3,1.5,1.8,2.0,2.3). The dotted lines represent models of constant metallicity
(\Zh=-0.71,-0.40,0.0,0.22). The metallicity and IMF slope values are quoted in
the left panel, where we show the grid corresponding to models of 12.6\,Gyr. In
the right panel we show the grid for   2\,Gyr. All the index measurements are
performed in the LIS-14.0\AA\ system.}
\label{fig:nadbcat}
\end{figure}

As it can be seen from Fig.~\ref{fratio} and Fig.~\ref{fig:nadb} this
feature strengthen with increasing IMF slope, whilst the Ca{\sc II} triplet at
$\sim$8600\,\AA\ weakens. This extent has been noted by \citet{Schiavon00} and
\citet{Conroy11}. In the latter it is shown that an increase in
{\mbox{$\mbox{[Na/Fe]}$}} causes a decrease in the strength of the Ca{\sc II}
triplet, as Na is a major contributor to the electron pressure in late-type
stars. In \citet{Vazdekis03} we fully characterize the behaviour of the Ca{\sc
II} triplet in SSPs. Therefore these two features can be used to propose an
index-index diagnostic diagram to constrain the IMF. Fig.~\ref{fig:nadbcat}
shows the resulting model grids for two different ages (2 and 12.6\,Gyr) for
models with Unimodal IMF. A range of IMF slopes and metallicities is shown. 
All the index measurements are performed on the LIS-14.0\AA\ system. These grids
allow us to distinguish the IMF slope if the age and the
{\mbox{$\mbox{[Na/Fe]}$}} abundance ratio are properly constrained.

\subsection{Galaxy spectrophotometric surveys}
\label{sec:surveys}

\begin{figure}
\includegraphics[angle=0,width=\columnwidth]{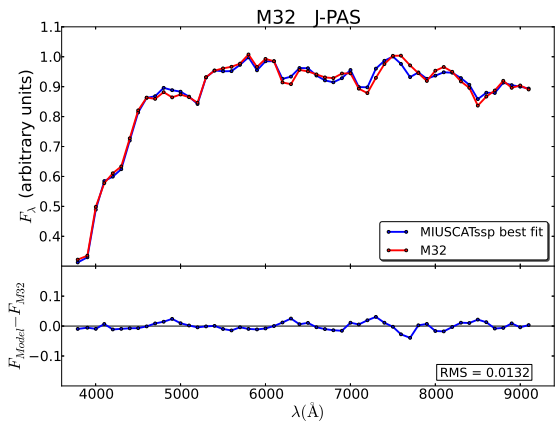}
\caption{ Spectral fitting of M\,32 using the MIUSCAT SSP SEDs. The
spectrum of the galaxy at the J-PAS resolution is plotted in red. The best
MIUSCAT SSP fit (Age = 4.6 $\pm$ 1.1\,Gyr; \Zh = 0.04 $\pm$ 0.08 dex) to the
spectrum of M\,32, as derived from a standard $\chi^2$ minimization technique, is
plotted in blue. The residuals are shown in the lower panel with the same
scale. See the text for more details.}
\label{fig:M32_J-PAS}
\end{figure} 

The use of multi-filter photometric surveys to determine SEDs and redshifts
with high enough level of accuracy (SDSS, see also COMBO-17: \citealt{Wolf08};
COSMOS: \citealt{Ilbert09}; ALHAMBRA: \citealt{Moles08}) has opened a new way
to analyze the stellar populations of galaxies at different z-values and in
different environments, thus allowing for the study of the evolution of
galaxies and cosmology using huge samples.

One of the major advantages of this kind of surveys is the fact that they
provide low resolution spectroscopy for every pixel of the sky. With low
resolution spectrophotometric data, spectral fitting techniques over the full
spectral range are mandatory to exploit as much as possible the information in
the data. A unique characteristic of this type of data is the fact that the
photo-spectra constructed on the basis of single narrow-band filter imaging
does not suffer from flux calibration systematic uncertainties, unlike standard
spectroscopy. Every single point of the photo-spectrum -i.e. every filter- is
observationally independent of the rest of the photo-spectrum, so the resulting
SED is not affected by large scale systematics in the relative flux calibration
(hence in the SED colors). With this unique advantage in mind, for the proper
analysis of the SEDs of galaxies and stars it is crucial to use template
stellar libraries with extremely accurate flux calibration, like MIUSCAT. In
this sense, MIUSCAT, and all the SED SSP models derived from this library, are
perfectly suited for the analysis and interpretation of optical
spectrophotometric data, provided the accurate flux calibration of the library
stars and the full optical spectral coverage. 

To illustrate a possible use of these new models, we focus on a test-case based
on the Javalambre-PAU Astrophysical Survey (J-PAS: \citealt{Benitez09};
\citealt{Cenarro10}), a new survey with 54 narrow-band (100\,\AA) contiguous
optical filters ($\lambda\lambda \sim$3500 -- 9500\,\AA) under preparation at
the Observatorio Astrof{\'{\i}}sico de Javalambre, which will provide low
resolution (R$\sim$30) spectrophotometric data for hundred million galaxies.
Figure \ref{fig:M32_J-PAS} illustrates the best SED fitting derived for  the
integrated spectrum of M\,32 from \citet{Bica90}, taken  from the compilation
of \citet{Santos02}, using the MIUSCAT SSP  models as input templates. M\,32
and the template spectra have  been convolved with the J-PAS filters to
simulate a real case. It is  clear from the figure that the best fit,
derived from a standard $\chi^2$ minimization technique, reproduces well the
observed spectrum at both  low and high frequencies. The obtained residuals are
shown in the lower panel. Note the telluric absorption still present in the
data in the spectral range $\lambda\lambda$ 7000--8000\,\AA.  The best solution
corresponds to a MIUSCAT SSP  model of 4.6 $\pm$ 1.1\,Gyr and around solar
metallicity (0.04 $\pm$ 0.08 dex),  in good agreement with results based on
much higher resolution  spectroscopic data (e.g. \citealt{VazdekisArimoto99};
\citealt{Schiavon04}).  

%%%%%%%%%%%%%%%%%%%%%%%%%%%%%%%%%%%%%%%%%%%%%%%%%%%%%%%%%%%%%%%%%%%%%%%%%%%%%%%%

\section{The web tool}
\label{sec:web}

The extension of the spectral range of our models based on the composite MIUSCAT
stellar spectra allows us to cover the gap left by our previous model
predictions based on the MILES and CaT libraries, as well as to extend blueward
and redward these libraries the wavelength coverage. Therefore this extension
represents a significant improvement that could be useful for many applications
as well as for model comparisons. To facilitate the exploitation of these models
we have integrated these new predictions within our recently developed website
(http://miles.iac.es). This web provides the necessary support for using these
model SEDs as well as those predictions based on MILES, CaT and Jones
(1999) stellar library.

All these models as well as the MILES and CaT stellar libraries can be retrieved
and handled according the requirements of the users.  The webtools include the
transformation of the spectra to match the instrumental set-up of the
observations (spectral resolution and sampling), measurements of line-strength
indices and synthetic magnitudes derived from the spectrum. 
Further descriptions of all these applications are described in Paper~II.

%%%%%%%%%%%%%%%%%%%%%%%%%%%%%%%%%%%%%%%%%%%%%%%%%%%%%%%%%%%%%%%%%%%%%%%%%%%%%%%%

\section{Summary and conclusions}

We have extended the spectral coverage of our stellar population synthesis
models by combining the predictions based on the MILES \citep{Sanchez06}, CaT
\citep{Cenarro01} and Indo-U.S. \citep{Valdes04} empirical stellar spectral
libraries. For this purpose we combined the stellar spectra of these three
libraries for a subsample of 432 stars of the Indo-U.S. database, with no gaps
in the relevant spectral regions and with no significant telluric absorption
residuals. However the whole Indo-U.S. stellar database was
taken into account for matching the stellar parameters of this library to the
MILES/CaT system. Due to the limited flux-calibration quality of the spectra of
the Indo-U.S. library, we used the MILES and CaT spectra as a reference. We
scaled these two spectral ranges according to empirical relations between the
stellar parameters and the Johnson-Cousin $V-I$ colour, which are based on
extensive photometric stellar libraries (mainly \citealt{Alonso96} for dwarfs and
\citealt{Alonso99} for giants). The same prescriptions were employed to convert
the theoretical parameters of the stellar isochrones \citep{Girardi00} that feed
the models to observable fluxes and colours. For the selected Indo-U.S. stars
that are lacking in either the MILES or CaT databases, or in the two, we
synthesized the corresponding stellar spectra for the same atmospheric
parameters using the MILES and CaT interpolators described in \citep{Vazdekis10}
and \citep{Vazdekis03}, which are also employed to generate the required spectra
to populate the isochrones during the SSP SED calculations.  

Appart of filling-in the spectral gap between the MILES and CaT libraries the
Indo-U.S. stellar spectra allowed us to slightly extend bluewards MILES and
redwards CaT the spectral ranges of these two libraries. Finally the spectral
ranges covered by the Indo-U.S. and CaT stellar libraries have been smoothed and
resampled to match the values of MILES. The resolution is therefore constant
along the whole spectral range, i.e. 2.5\,\AA\ (FWHM), with 0.9\,\AA\ per
pixel. The spectral coverage of the composite stellar spectral range is
$\lambda\lambda$ 3464.9 -- 9468.8\,\AA.

The composite MIUSCAT stellar spectra were implemented in our models to compute
SSP SEDs based on these spectra. These models were then combined with the
published model SEDs for the MILES \citep{Vazdekis10} and CaT \citep{Vazdekis03}
following a similar approach to that employed for the MIUSCAT stars, keeping the
SSP flux scaling for these two spectral regions. Therefore in the composite
MIUSCAT SSPs the SEDs in the MILES and CaT spectral ranges are identical to the
originally published models, with the exception of the resolution of the CaT
models, which have been smoothed to match that of MILES. We estimated the
reliability of the resulting models according to the method described in
\citet{Vazdekis10}, which takes into account the atmospheric parameter coverage
of the 432 stars included in the MIUSCAT stellar library. The new model SEDs are
reliable in the metallicity range ($-0.71 \leq \Zh \leq 0.22$) for all the ages
(0.063 -- 17.8\,Gyr) and IMF shapes (Unimodal, Bimodal, Kroupa Universal and
Kroupa Revised, as described in \citet{Vazdekis03}) and IMF slopes
(0.3$\leq$$\mu$$\leq$2.3) that are considered in this work. 
Broad band colours are also acceptable outside these ranges.

The new models allow us to measure colours involving the Johnson-Cousins and
SDSS broad-band filters from $U$ to $I$ and from $u$ to $i$, respectively. To
measure the magnitude in the $U$ filter we compute a correction factor for the
missing flux, as this filter extends bluewards the MIUSCAT spectral range. This
correction is computed with the aid of the \citet{Pickles98} low-resolution
stellar library. The good flux-calibration quality of the resulting model
spectra allows us to measure accurate colours in the covered spectral range. The
derived SSP colours are in very good agreement with the corresponding
predictions obtained on the basis of empirical relations between colours and
stellar parameters, which come from extensive photometric libraries. 

These model SEDs can be particularly useful for galaxy surveys, allowing to
study the behaviour of the model SEDs with relevant stellar population
parameters for a variety of filter responses.

Unlike models based on theoretical stellar atmospheres our SSP models provide
good fits to the colours of the globular clusters of the Milky Way for the ages
and metallicities that are derived from CMD studies. More extensive comparisons
with globular cluster data and galaxies are shown in the  second paper of this
series (Paper~II). We show the dependence of the colours as a function of age,
metallicity and IMF. We find that colours involving redder filters show greater
sensitivities to the IMF slope.

We also find various absorption line features and molecular bands that are very
sensitive to the IMF slope throughout the whole spectral range covered by the
MIUSCAT model SEDs. Among the most prominent features we find that the Na{\sc
I} doublet at $\sim$8200\,\AA\ increases its strength with increasing IMF
slope, whereas the neighboring Ca{\sc II} triplet around $\sim$8600\,\AA,
decreases, confirming previous findings (e.g., \citealt{Schiavon00};
\citealt{Vazdekis03}; \citealt{Conroy11}). Note that the Na{\sc I} feature is
not covered by the model SEDs based on the MILES and CaT stellar spectral
libraries. 

 With the aid of the MIUSCAT SSP SEDs we propose a new index definition for the
Na{\sc I} doublet at $\sim$8200\,\AA, Na{\sc I}8200A, which has significant
advantages over previous index definitions for this feature, particularly for
low velocity dispersion stellar systems. Na{\sc I}8200A varies with the age and
the metallicity and it is particularly sensitive to the IMF slope. We
propose a Na{\sc I}8200A -- CaT$^*$ diagram, the latter index defined in
\citet{Cenarro01}, to distinguish the effects of the IMF slope if the age and the
{\mbox{$\mbox{[Na/Fe]}$}} abundance are properly constrained.

Finally we show the utility of these new models for survey-oriented
spectrophotometric applications. Because of the wide spectral coverage 
and accurate flux calibration of the MIUSCAT library stars, the SSP models 
are perfectly suited for being used as templates for exploiting galactic
spectrophotometric data, deriving reliable estimations of ages and metallicities 
even for low resolution data. 

The MIUSCAT model SEDs, as well as all our model predictions, can be downloaded
from the miles website (http://miles.iac.es), where web-based facilities are
available for downloading and handling these SEDs. These user-friendly tools also
allow to obtain the MIUSCAT SEDs, with the corresponding line-strength indices
and colours, for a variety of SFH parametrizations as well as for user-defined
SFHs.

%%%%%%%%%%%%%%%%%%%%%%%%%%%%%%%%%%%%%%%%%%%%%%%%%%%%%%%%%%%%%%%%%%%%%%%%%%%%%%%%

\section*{Acknowledgments}

We would like to thank the Indo-U.S. team for making available their stellar
spectral library. We thank F. Valdes and A. Alonso for very useful
clarifications on their respective libraries as well as E. M\'armol-Queralt\'o
for testing a preliminary version of the MIUSCAT SEDs that allowed us to improve
the models. We also thank the referee for relevant suggestions that certainly
improved the original draft. The MILES and CaT libraries were observed at the
INT and JKT telescopes, respectively, on the island of La Palma, operated by the
Isaac Newton Group at the Observatorio del Roque de los Muchachos of the
Instituto de Astrof\'{\i}sica de Canarias. This research has made an extensive
use of the SIMBAD data base (operated at CDS, Strasbourg, France), the NASA's
Astrophysics Data System Article Service, and the {\it Hipparcos} Input
Catalogue. AJC and JFB acknowledges support from the Ram\'on y Cajal Program
financed by the Spanish Ministry of Science and Innovation. AJC and JFB are {\it
Ram\'on y Cajal} Fellows of the Spanish Ministry of Science and Innovation. This
work has been supported by the Programa Nacional de Astronom\'{\i}a y
Astrof\'{\i}sica of the Spanish Ministry of Science and Innovation under grants
AYA2010- 21322-C03-01 and AYA2010-21322-C03-02 and by the Generalitat Valenciana
under grant PROMETEO-2009-103.

%%%%%%%%%%%%%%%%%%%%%%%%%%%%%%%%%%%%%%%%%%%%%%%%%%%%%%%%%%%%%%%%%%%%%%%%%%%%%%%%

\label{lastpage}
\end{document}